\documentclass[runningheads,a4paper,orivec]{llncs} 
\usepackage{amssymb} 
  \usepackage{xcolor} 
  \usepackage{transparent}
\usepackage{graphicx}
\usepackage{amsmath}
\usepackage[numbers]{natbib}
\graphicspath{{}} 
\title{Modular programming of computing media using spatial types,  for   artificial physics.
 }\titlerunning{ Modular programming of computing media}  
 \author{Fr\'ed\'eric Gruau }
 \institute{Laboratoire de Recherche en Informatique, Universit\'e paris 11,  Orsay, France  }
\begin{document}   \maketitle     
   
  \begin{abstract}
  Our long term goal is to execute General Purpose   computation on homogeneous computing media consisting of  millions of small identical Processing Elements (PE) communicating locally. We proceed by simulating the Self-Development of a Network (SDN) of membranes, and this implies a medium able to implement artificial physics laws that simulates  simplified membrane-agents, dividing and homogenizing. This is a difficult challenge: our current  version of SDN-media uses  PEs with 77 bits of state and 13878 gates.
   This high level of complexity forced us to work out 
   an efficient and expressive  scheme for programming the medium, the goal of this paper it to present it.
   
   The PE's communication graph has to be a maximal planar graph.
   Fields of bits are spread in 2D, over three locus: the vertices, edges and faces of this planar graph.  They constitute three data types, which abstract away the ensemble of PEs. 
   The simplicial proximity between bits  is used to define operations on fields, thus implementing    ``{\it spatial type}". Expression combining  operations can be translated in logical circuits . 
   
The efficiency is achieved because fields of different locus are combined using reduction  operation. This allows to factorize computation by exploiting  the symmetries always present when simulating physics.

 The expressiveness is achieved  by allowing a modular procedural  programming: Instead of directly focusing on a specific target update function, we develop a library or reusable functions mapping fields to other fields.

For illustrating efficiency and expressiveness, we choose to program  a key  building  block of the SDN-medium, and reuse it for  generating a logical circuit computing the discrete Vorono\"{i} diagram  above the  planar graph. 
       
  We consider two kinds of maximal planar graph: with isotropic distribution of PEs or with the hexagonal lattice structure. The first compares to amorphous   computing medium and has a better potential for hardware scalability, the second compares  with  cellular automata computing medium, it is more efficient. 
\end{abstract}

\section{Introduction}

\subsection{ Physics on Computing Media for general-purpose Computing.}

\paragraph{Computing Media.}
 Future computing platforms, whether very-large-scale integration (VLSI), nano, or bio ,  will probably consist of a vast number of Processing Elements (PEs) homogeneously embedded in 2D or 3D space, where the magnitude involved forces the programmer to incorporate the
\textit{locality constraint}: Each PE has a specific location in space, communication is local in space,    communication time must be proportional to  Euclidian distance as in the VLSI complexity model~\cite{vlsiComplexity}. 
This invariant enables unbounded scalability of hardware, and characterizes a  family of computer architectures refereed to as  \textit{``computing  media''}~\cite{programming-blob}. :
 %While communication costs have always been a major issue in parallel computing, scaling up to an arbitrary large space is rarely considered as an option, where communication must be optimized by taking into account physical distance. 
   This family includes  \textit{regular} classic models, such as  Cellular Automata (CA) or systolic arrays; but also \textit{irregular} models where the constraints of lattice
tiling of space and synchronism in time are relaxed, as exemplified in the amorphous medium~\cite{amorphous}  
% In general, what makes the whole GPCA-project   worthwhile, is that it could deliver a general-purpose parallel architecture with potentially unlimited power, as $n$ grows unbounded. 
%   In this context, it is relevant to consider the amorphous computing medium wich is a more scalable medium than CA, because it relax the  requirement of a lattice, and global synchronisation.
% \paragraph{Relaxing the lattice requirement.}
which  is an homogeneous and isotropic  scattering of  PEs in 2D or 3D space,   with nearest-neighbor communication.

\paragraph{Physics on Computing Media.} Because they are tightly bound to space, simulating physics is what computing media are naturally good at. It is a major   application  of CAs. Physical laws expressed as differential equations can be translated into simple local rules~\cite{chopard}. Physics can also be done without lattice discretization of space:
      Rauch~\cite{rauch}   modeled    wave   propagation on an amorphous medium.

\paragraph{Towards General Purpose computing media.}  Simulating physics is only a tiny fraction of the spectrum of computation as we know it. While computing media have a potentially unbounded hardware scalability, they cover   a very narrow scope of application. Our long term project is to broaden this scope up to general purpose  computation. 
To achieve this level of general, one more level of indirection is necessary. We implement a virtual machine on top of the computing medium, called {\it self-developing network}~\cite{SDG,SDM}.    
%This layer offer higher level programming primitives allowing to program a computing medium in a more abstract way:  In short, we use physics to go beyond physics! 

\paragraph{Self Developing Network(SDN).}  The SDN virtual machine allows to program ``space agnostic'' real algorithms. In~\cite{programming-blob} we showed how to program and execute  matrix multiplication and sorting.
Simulating this machine means implementing  on a medium physical objects which are biological artifacts:  simplified membranes modeled as connected blobs.   Membranes allows to structure space into independent regions where distinct computation can take place. This is ``artificial physics'': the goal is not to model reality  but to design an SDN-medium emulating  the SDN virtual machine. In other words,   using physics to go beyond physics.
For example we set repulsive forces between membrane to homogenize their distribution (load balancing), and strangle force to divide membranes (self-development).  %I  
%  We engineered the following building blocks: \begin{enumerate}
%    \item   Represent simplified membranes   separating   inside from  outside. 
%    \item   Homogenize their placement  through computation of repulsive forces.
%    \item   Divide membranes using another strangle force, triggered by host-instructions. 
%  \end{enumerate}
  Execution  resembles a much simplified biological developmental-process. What is developed is not a multicellular organism, but  a clean and deterministic virtual network of virtual processing-elements delimited by membrane-blob and adherence between them.  The   connectivity is determined by the instructions. 

\paragraph{The current status of the SDN-medium.}  
Our current version  can interpret a flow of host-instructions dictating  the  self-development of a virtual 2D-grid network of membrane, in a time proportional to the diameter of the circuit. It is shown in  short youtube videos in~\cite{utubeDevelopement2018}. This  demonstrates that  efficient general-purpose execution on computing media is not an utopia.

\paragraph{Cellular Circuit, amorphous computers and Cellular Automata.} The SDN-medium is quite complex.
 Its programmation and simulation was made possible thanks to a new scheme which allows a modular specification and an efficient execution. It is based on spatial types, which
 embed data and operation in 2D space.
   The  first  goal of this paper is to explain spatial types, and why it enables to tackle complexity thanks to  improved efficiency and modularity.
  A program using spatial types is translated into   a circuit of logic gates   embedded in 2D space.   Just like CAs, the same computation goes on through space justifying the denomination ``cellular circuits''; However, as in amorphous computers, the circuit's global structure is not constrained to be a lattice. 
   If  a lattice is used, though, simulation is much more efficient, and cellular circuits becomes CAs. We will now contrast cellular circuits with respect to amorphous computing, and  CAs.

 \begin{figure} \centering \def\svgwidth{\columnwidth}  
%% Accompanies image file 'twoPlanarGraph.pdf' (pdf, eps, ps)
\begingroup%
  \makeatletter%
  \providecommand\color[2][]{%
    \errmessage{(Inkscape) Color is used for the text in Inkscape, but the package 'color.sty' is not loaded}%
    \renewcommand\color[2][]{}%
  }%
  \providecommand\transparent[1]{%
    \errmessage{(Inkscape) Transparency is used (non-zero) for the text in Inkscape, but the package 'transparent.sty' is not loaded}%
    \renewcommand\transparent[1]{}%
  }%
  \providecommand\rotatebox[2]{#2}%
  \newcommand*\fsize{\dimexpr\f@size pt\relax}%
  \newcommand*\lineheight[1]{\fontsize{\fsize}{#1\fsize}\selectfont}%
  \ifx\svgwidth\undefined%
    \setlength{\unitlength}{422.95730577bp}%
    \ifx\svgscale\undefined%
      \relax%
    \else%
      \setlength{\unitlength}{\unitlength * \real{\svgscale}}%
    \fi%
  \else%
    \setlength{\unitlength}{\svgwidth}%
  \fi%
  \global\let\svgwidth\undefined%
  \global\let\svgscale\undefined%
  \makeatother%
  \begin{picture}(1,0.38068796)%
    \lineheight{1}%
    \setlength\tabcolsep{0pt}%
    \put(0,0){\includegraphics[width=\unitlength,page=1]{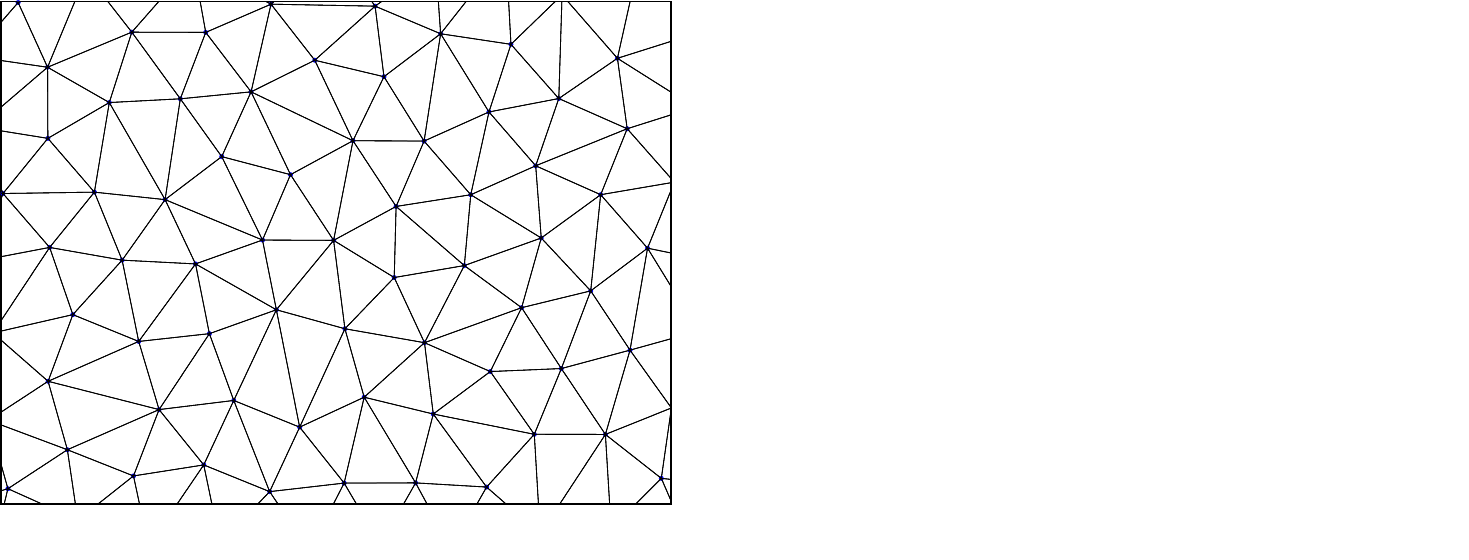}}%
    \put(0.19053246,0.0037404){\color[rgb]{0,0,0}\makebox(0,0)[lt]{\lineheight{1.25}\smash{\begin{tabular}[t]{l}(a)\end{tabular}}}}%
    \put(0.74745382,0.0037404){\color[rgb]{0,0,0}\makebox(0,0)[lt]{\lineheight{1.25}\smash{\begin{tabular}[t]{l}(b)\end{tabular}}}}%
    \put(0,0){\includegraphics[width=\unitlength,page=2]{twoPlanarGraph.pdf}}%
  \end{picture}%
\endgroup%

\caption{Two target architectures for the underlying maximal planar graph linking PEs (a) homogenous isotropic (b)   hexagonal lattice.
% equivalent to  Cellular Automata. 
} \label{fig:twoPlanarGraph} \end{figure}
 
 \subsection{Isotropic  cellular circuit and amorphous computing}
A basic amorphous medium is   rough: PEs do not know their spatial location, they receive a radio signal sent by nearby neighbors.  
Research has focused on computing low level information  such as a simple pair of approximate 2D coordinate~\cite{amorphousCoordinate}\footnote{ Coore~\cite{coore} proposed a more generic technique having the potential of installing arbitrary patterns of blobs, with a target topological arrangement. However, only small patterns where demonstrated.}.
In contrast, cellular circuits   really enable  the programming of complex behavior~\cite{utubeDevelopement2018}.
However, this is achieved at the cost of considering  a ``cleaner'' medium endowed with a specific property: the PEs have to communicate between themselves following a network which  must be a maximal planar graph (all the faces are triangles). 
It can be seen as a preliminary layer which has to be installed on a basic  amorphous medium,   enhancing its programmability.  We know of two possible solutions: 
1- If the PEs know their 2D coordinates,  doing a Delaunay triangulation   directly builds  a maximal planar graph. 2- Otherwise   a ``{\it combinatorial Delaunay graph}'' can be  built  using only the hop counts between PEs. In~\cite{distributedDelaunay}, only a subset of the   PE in a sensor network are linked, the subset is computed iteratively so as to create a  Centroid Voronoi Tessellation, which means in short, an homogeneous distribution as close as possible to the hexagonal lattice. 
 
 \paragraph{Synchronous versus asynchronous} In contrast to amorphous medium
  we simulate a synchronous framework of cycles, during which each PEs receives all the messages of its neighbor and then updates its state. However, with a small uniform probability we choose  not to  update the state. This is a significant step towards  the asynchronous framework. It generates random fluctuation in the SDN medium, and is in fact beneficial: it is a natural   way to solve conflicts.
  
 \paragraph{Homogeneous and isotropic distribution.} For amorphous media, a simple and often used PE distribution in 2D is  {\it ``Poisson-disk''} sampling: the  location of each PE is chosen with a uniform probability, but discarded if there are already other  PEs nearby (within a disk of a given radius). We used  the Furthest Point Optimization (FPO) algorithm~\cite{FPO} which produces more homogeneous and isotropic distribution. 
 The resulting planar graph  (produced with Delaunay triangulation) is shown in fig.~\ref{fig:twoPlanarGraph}~(a).  The  improved quality causes the hop-count distance to become  a good approximation of the geometric distance, and this in turns, enables to compute spatial features with accuracy. In the example of this paper, we  will see that the  hop-count discrete Vorono\"{i} Diagram (VD) approximates the real  VD.

\subsection{Contribution of hexagonal cellular circuits to CA.}
 The hexagonal lattice shown in fig.~\ref{fig:twoPlanarGraph}~(b) is a non-isotropic (6 obvious preferred directions), but very regular maximal planar graph.
 This paper is an extended version of~\cite{GruauTPNC}, which considered only the hexagonal network option.
 In this option, a cellular circuit can be translated into a CA, so spatial types can be understood as a different scheme for specifying CAs. What are the advantages?
 
%, the radius is incremented at each communication.
% this explains   why the SDN-medium reaches a radius as high as 25. 
 
% We will also use it for illustrations because it is easier to follow. 

% For the GP-medium, the equivalent CA-cell uses 77 bits of memory, and   the next state  needs a circuit of 13878 gates, and  a radius of 25. 

\paragraph{Specifying rotation-invariant rule in a more expressive way.} The next state of a CA cell is programmed as a function of its neighborhood.
In contrast, cellular circuits are inspired from *Lisp~\cite{hillis1989connection}, considering fields.
%which was used to program the connection machine    
 Boolean fields are  defined on sets of points in  2D space called  ``{\it locus}''.  %defined over the whole cellular space, 
Fields are combined using  reductions which apply a commutative associative operation on values  found in the immediate spatial  neighborhood of each point.  One great virtue of reduction is that the order in which the neighbors are processed does not matter. In particular, neighbors need not be distinguished, and this ensures rotation invariance. Totalistic CA~\cite{Wolfram1983}, which sums   the immediate neighbor's state is a CA-illustration of this  principle.
With cellular circuits, we use any type of reductions, not just the sum.
More fundamentally, by carefully constructing 9 locus,  we are able to define 12 different type of  neighborhood on which to reduce.
% % the neighbor   by first   making a $+$ reduction, (sum)  of their  state:%, which makes it  also rotation invariant.  
%  \begin{enumerate}
%    \item   Any commutative-associative operations is used (not just addition).
%   \item Fields are defined  on vertices (V)   Edges (E) and Faces (F)  of a  planar graph
%  \end{enumerate}
%   %  and points in-between: eV, fV, vE, fE, vF,eF.  
%   The simplicial adjacency between   those three class of V,E,F bits  provides twelve different neighborhood on which to reduce. 
This   increases expressiveness to the point that   being forced to compute using reductions is not experienced as a constraint. On the contrary, it feels just natural.
   When doing physics, we always do compute rotation-invariant quantities anyway, so embedding rotation-invariance in the operations themselves incorporates  a useful domain-specific information which %significantly 
   alleviates the task.

\paragraph{Generating and simulating rules with high radius.} An hexagonal cellular circuit can be  translated into a classic hexagonal CA, however this is mainly a theoretical statement. 
Usually, a CA next-state update rule consults only the immediate neighborhood, which is  radius 1.  
Considering bigger neighborhood with higher radius is not natural, the idea being to keep the rule simple.
With spatial types, we do specify update rules which process large neighborhood, but at the same time also remaining very simple. For example, a simple sequential composition of $r$ reductions produces a field of radius $r$. This is because  each time a reduction is applied, it creates a communication thus incrementing the radius of the resulting field. 
 %In practice, using only reductions  allows to factorize computation,  by exploiting symmetries.
  As a result, fields of radius $r$  can be computed  in time $O(r)$,  which   becomes $O(r^3)$ if translated in a CA. 
  In practice, when using spatial types, fields having high radius is the normality not the exception. In other words, spatial types allow to explore and simulate a different portion of the landscape of CAs.  For the SDN-medium, we compute fields of radius up to $r=25$, which render the interactive simulation of a  CA-translation  unfeasible.

\paragraph{Modularity.} This is the most important property. By modularity we mean encapsulating code into   functions  that can be reused,  several time within the same CA, or from one CA to another distinct CA: We illustrate both cases in this paper.
Modularity is obtained because we compute fields. We can compute them as the result of a function call taking other fields in parameter. Functions of generic interest naturally pop up. By composing them, one obtains quickly very complex behavior. 

% Using operation-expression on fields, one does not directly focus on the final update function as in CAs. Instead, one proceeds by  debugging a generic library of  functions mapping fields to other fields. Those functions are defined once, but can be called several times with different parameters.
% %, and this is what we mean by a modular specification. 
% For example the function $x\mapsto$meet$(x)$ programmed in  section~\ref{sec:meet}   is a key function used  many time in the SDN-medium, for constraining agent's moves, so as to avoid merging or dividing their supports. Subsection~\ref{subsec:VD} uses a part of $x\mapsto$meet$(x)$, in order to compute the discrete Vorono\"{i} diagram. 
%  % operation-expression thus allows  a  modular-programming.
%  

%  We  implement reduction of a boolean field  on vertices  (resp. Edge, Face)  to generate another boolean field on F or E (resp. F or V, V or E). This is done using six intermediate locus called t-locus, eV,vE,fV,vF, fE,eF inserted in between (V,E), (V,F) (E,F). 
% T-locus allows a clean decomposition of this first six primary reduction.
%  and also the definition of six other secondary reduction between fields on t-locus.
% If the planar graph is homogenous and isotropic, then, 
% the operations will capture the uniformity of space.  

%  
%  \begin{figure} \centering \def\svgwidth{\columnwidth}  
%  \includesvg{twoGraphPoint}
% \caption{Mapping the VEF data points, in 2D. (a) homogenous isotropic (b) the hexagonal lattice.
% % equivalent to  Cellular Automata. 
% } \label{fig:twoGraphPoint} \end{figure}
%  

\section{A 2D spatial type based on maximal  planar graph.} 

Informally,  a ``{\it 2D spatial type}'' is a set of data embedded  data  in 2D space, and proximity is used to define operation. 
  Instead of using a lattice to define location and proximity, as is done in CA, we will need only a planar graph, and use its faces and edges to locate data in 2D.

\paragraph{The simplicial graph.} A connected graph which can be drawn without any edges crossing, is called planar. When a planar graph $G$ is drawn in this way, it divides the plane into regions called faces.  
A planar graph $G$ can be represented very naturally by another graph $G_S$~\cite{grigor2014graphs}, called ``{\it simplicial graph}'':
The  set  of  vertices  of $G_S$
coincides  with  the  set  of  all  simplexes $S$ of $G$, which consist of  three classes: vertices of dimension 0, edges of dimension 1, and faces of dimension 2. 
 Two simplexes $s,t$ are connected in $G_S$ if and only if $s\subset t$ or $t \subset s$ in $S$. In other words, a vertices (resp an edge) is adjacent to the edges, and faces (resp. to the two  faces) including it.

\paragraph{Maximal planar graph.} It is a planar graph where no edges can be added without breaking the planarity, which implies that all the faces are triangles.
 We will consider exclusively maximal planar graph, because this property is needed for defining some of the operations.
 Let the vertex count be $V$, edge count $E$, and face count $F$.
Maximal planar graph verify  $2E=3F$, 	as can be derived by taking the sum over every face of the number of edges in each face which is 3. We also have $V-E+F=2$, (Euler's formula) hence  $F=2V-4,E=3V-6$.
The arity is a number associated to each class of simplex; it represents the proportion of the simplexes in each classes: It is 1 for vertices, 2 for faces, and 3 for edges.

\paragraph{The V,E,F simplicial locus.} From an embedding in 2D of a maximal planar graph $G$ where edges are drawn with straight lines,   the simplicial graph $G_S$ can be also embedded in 2D as another planar graph, by locating its vertices: %as shown in fig.~\ref{fig:twoGraphPoint}):
  The vertex-vertices map to the vertices of $G$,  the edge-vertices to the edge's middle,
    and the face-vertices to the face 's barycenter. 
  Bits of data will be conceptually associated to those 2D points  hence we call them {\it ``data-points''}. Those three set of data-points are called  respectively the V,E,F locus. We refer to them as the {\it ``simplicial locus''},   so as to distinguish them from other locus introduced later.

\begin{figure} \centering \def\svgwidth{\columnwidth}  
%% Accompanies image file 'VEFtiling.pdf' (pdf, eps, ps)
\begingroup%
  \makeatletter%
  \providecommand\color[2][]{%
    \errmessage{(Inkscape) Color is used for the text in Inkscape, but the package 'color.sty' is not loaded}%
    \renewcommand\color[2][]{}%
  }%
  \providecommand\transparent[1]{%
    \errmessage{(Inkscape) Transparency is used (non-zero) for the text in Inkscape, but the package 'transparent.sty' is not loaded}%
    \renewcommand\transparent[1]{}%
  }%
  \providecommand\rotatebox[2]{#2}%
  \newcommand*\fsize{\dimexpr\f@size pt\relax}%
  \newcommand*\lineheight[1]{\fontsize{\fsize}{#1\fsize}\selectfont}%
  \ifx\svgwidth\undefined%
    \setlength{\unitlength}{356.17389769bp}%
    \ifx\svgscale\undefined%
      \relax%
    \else%
      \setlength{\unitlength}{\unitlength * \real{\svgscale}}%
    \fi%
  \else%
    \setlength{\unitlength}{\svgwidth}%
  \fi%
  \global\let\svgwidth\undefined%
  \global\let\svgscale\undefined%
  \makeatother%
  \begin{picture}(1,0.41674609)%
    \lineheight{1}%
    \setlength\tabcolsep{0pt}%
    \put(0,0){\includegraphics[width=\unitlength,page=1]{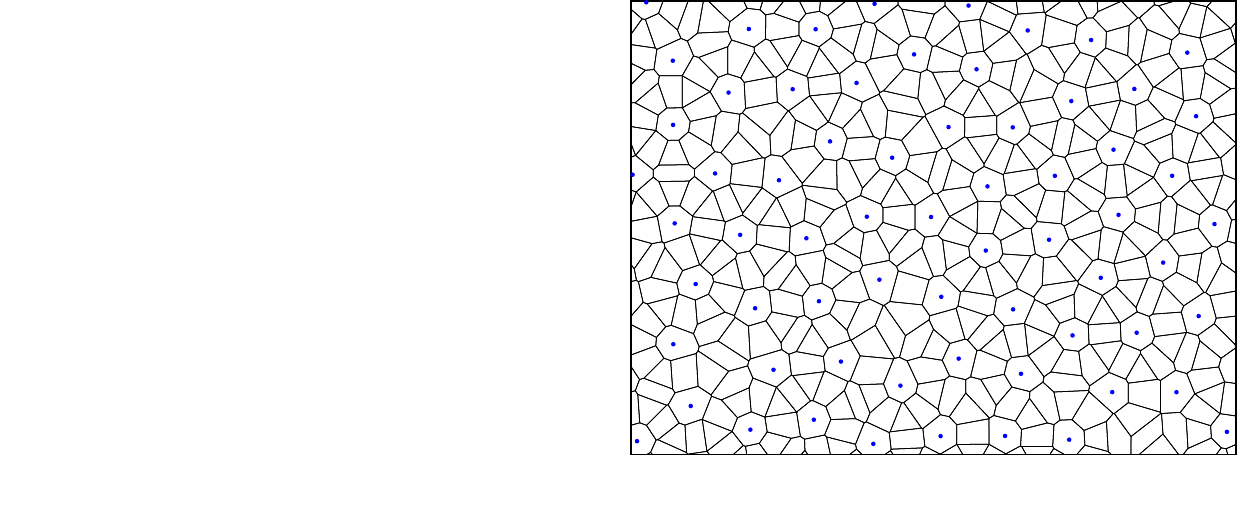}}%
    \put(0.72307433,0.01071435){\color[rgb]{0,0,0}\makebox(0,0)[lt]{\lineheight{0}\smash{\begin{tabular}[t]{l}(b)\end{tabular}}}}%
    \put(0,0){\includegraphics[width=\unitlength,page=2]{VEFtiling.pdf}}%
    \put(0.21555725,0.00407159){\color[rgb]{0,0,0}\makebox(0,0)[lt]{\lineheight{0}\smash{\begin{tabular}[t]{l}(a)\end{tabular}}}}%
  \end{picture}%
\endgroup%

\caption{The VEF tiling associated to (a) the hexagonal lattice  (b) the homogeneous isotropic planar  graph} \label{fig:VEFtiling} \end{figure}
\subsection{Computing blob features by reducing  simplicial fields.}
   
\paragraph{Simplicial fields.} Spatial types are boolean fields, more precisely: function from one locus to  $\{0,1 \}$.   The  type  is called  boolV (resp.   boolE, boolF), for Vertices (resp. Edge, Faces).  We also use integer fields  with a   small number of bits, usually 2 or 3.  For example, for two bits, the type is noted  int2V, int2E, int2F.
 The bit density of a field is this number   multiplied  by the  arity of the simplex: int2E costs 2*3=6 bits. The total memory needed for a field is its bit density, multiplied by the number of vertices,   minus a small constant.
 
\begin{figure} \centering \def\svgwidth{\columnwidth}  
  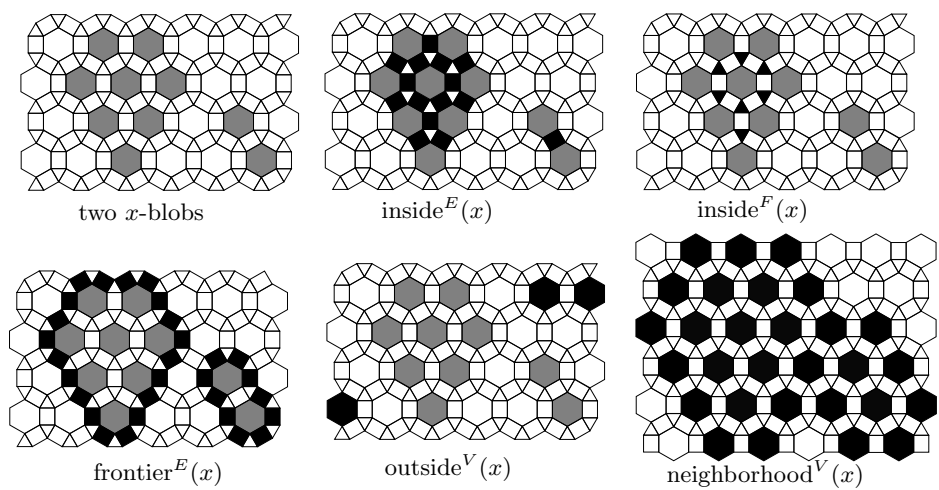
\caption{Boolean Fields (black) encoding  features of  two  $x$-blobs (gray).
%, on the hexagonal lattice.
 A boolV (resp. boolE, boolF )is  a  set of hexagons (resp. rectangles, triangles). % and is represented  by coloring that set. 
From  a boolV $x$ representing two  $x$-blobs, generic featuresof $x$-blobs can be computed from $x$ using simplicial reductions.
%resulting from one reduction for inside$^E$,inside$^F$, border$^F$, a composition for outside$^V$ neighborhood$^V$.
} \label{fig:feature} \end{figure}

 \paragraph{Representation of simplicial fields.}
 we draw the Vorono\"{i} Diagram of the three VEF   locus taken together, shown in fig.~\ref{fig:VEFtiling}. 
   Fig.~\ref{fig:VEFtiling} shows  the tiling obtained for the two planar graphs of fig.~\ref{fig:twoPlanarGraph}: For the hexagonal lattice in fig.~\ref{fig:VEFtiling}~(a), the tile of vertices, (resp. edges, faces) are hexagons, (resp. rectangles, triangles).   
This tiling is known as the "{\it Rhombitrihexagonal}" tiling. It is a  beautiful   Archimedian tiling used in architecture. 
 In order to represent a field, we  color the   tiles of the   subset  of data-points   for which the field is true.  For this reason, false and true are often called ``empty'' and ''filled''. 
 Fig.~\ref{fig:feature} shows some example of boolV,boolE and boolF using the Rhombitrihexagonal tiling.  
  %For the amorphous version in fig.~\ref{fig:VEFtiling}~(b), edges and face 's tile also resemble rectangle and triangle. The  vertices's tile is most often an hexagon, but sometimes it has 5 or 7 sides. 
 % We will use the Archimedian tiling for illustration because it is simpler. The amorphous tiling will be used to run the VD circuit.

  \paragraph{Six simplicial reduction between simplicial locus.} 
  Let $X,Y\in \{ $V,E,F$ \}$ be two distinct simplicial locus. From a bool$X$, one can compute a bool$Y$, as follows: For each point $p$ of the target locus $Y$, we apply a bit reduction  AND, (resp. OR, XOR) of all the point in locus $X$ which are simplicial neighbor  of $p$. We call the corresponding operation $\forall^Y$ (resp. $ \exists^Y, \delta^Y$). The upper script indicates the target locus $Y$. The he number of neighbors of $p$ is called the ``{\it co-arity}''. In the hexagonal case,  for the three simplicial locus, we have arity$*$co-arity $=6$.
  co-arity $-1$ is the number of binary gates needed to do the reduction. We must multiply by the arity of $Y$ to obtain the gate density which is therefore $\mathrm{arity}(Y)*(6/\mathrm{arity}(Y)-1)$. For $Y$=V (resp. E, F) it is 5 resp (3, 4).
   A ``{\it simplicial reduction}'' is overloaded since it can be applied to any of  the two other locus, for example $\forall^E$ produces a bool$E$ from either a bool$V$ or  a bool$F$.
  Simplicial reductions also exist for integer field, for example with min, max, or plus.  A simplicial reduction is a ``{\it spatial operations}'', because it  uses proximity in space.  
  We also use non-spatial operations  applying an operation separately on each data-point.
  For example $x\mapsto \neg x$ can be applied on a bool$V$ (resp. a bool$E$,   a bool$F$) to produce  a new bool$V$ (resp. bool$E$,  bool$F$). In this case, the gate density is the arity. %It compiles into a simple not gate, on the corresponding locus, so the gate density is the arity of the locus. 

 \paragraph{Computing blob features using simplicial reductions.} An SDN-medium   uses non-punctual agents whose support spans a set of vertices. It is thus represented using a bool$V$. Supports are separated by considering connected components for vertex-adjacency: 
\begin{definition} Let $x$ be a boolV.  $x$-blobs (resp. $x$-holes) are connected components of filled (resp. empty) vertices. \end{definition}  
 Arbitrary many $x$-blobs can be encoded with a single bool$V$ $x$, provided there is enough space. Using simplicial reductions, we can easily compute simple 2D-features of $x$-blobs, shown in fig.~\ref{fig:feature}.
%If two overlaping set are drawn, colors are added. Colors gets automatically allocated so that is adding is possible.
\begin{itemize}
  \item Function $x\mapsto$ frontier$^E(x)=\delta^E(x)$ (resp.   inside$^E(x)=\forall^E(x)$,  outside$^E(x)=\forall^E(\neg x)$ ) is the set of edges   adjacent to both an empty and filled (resp.  to only filled, to only empty) vertices. It costs 3 (resp. 3, 4) gates.
 Function $x\mapsto$  inside$^F(x)=\forall^F(x)$  is the set of face   adjacent to only filled vertices, it costs 4 gates. 
\item  Function $x\mapsto$  inside$^V(x)=\forall^V($inside$^E(x))$ (resp.   outside$^V(x)=\forall^V($outside$^E(x))$ is the set of filled (resp. empty) vertices surounded by  vertices in the same blob (resp. hole). It costs 3+5=8 (resp. 4+5=9) gates.
%\item  The   field  frontier$^V(x)=\exists^V($frontier$^E(x))$  is the set of vertices adjacent to  frontier$^E(x)$. It is partitioned into:  inFrontier$^V(x)=$frontier$^V(x)\wedge x$ and   outFrontier$^V(x)=$frontier$^V(x)\wedge \neg x$. They cost 8,9,10 gates.
\item  Function $x\mapsto$  neighborhood$^V(x)= \exists^V(\exists^E(x))$  costs 3+5=8 gates.
\end{itemize}

\paragraph{The radius of an operation expression. }
  CAs uses the notion of ``radius''  of the neighborhood to consider for computing the next state. It is an important concept which is also defined for spatial types, though at a finer granularity than vertices:
 \begin{definition}
The radius of a function is  the max distance ( hop-count between V,E,F  locus) to data-points of parameters influencing the result.
 \end{definition}
% The radius is incremented when a transfer operation $\uparrow^V, \uparrow^E, \uparrow^F$ is applied.
For the preceding small functions, which are all taking a bool$V$ as input, 
the radius is 1 for the bool$E$, and bool$F$ functions, and 2 for bool$V$ functions. Indeed, going from one vertex    to the neighbor vertex  takes two hops.%, since one needs to go through an edge.

% 
% We consider a set of PEs in the 2D plane, its Vorono\"{i}d Diagram and the dual Delaunay Tesselation.  
% Our purpose, stated in the introduction, is to design an architecture achieving  both scalability, and general purpose.
% CA are scalable because of locality of connections, but there exists more scalable architecture: namely, amorphous computer, which relax the requirements of christal regularity, and 
% simulate efficiently cellular computation doing artificial physics.

 \begin{figure}\centering   \def\svgwidth{\columnwidth}  
%% Accompanies image file 'tLocusTile.pdf' (pdf, eps, ps)
\begingroup%
  \makeatletter%
  \providecommand\color[2][]{%
    \errmessage{(Inkscape) Color is used for the text in Inkscape, but the package 'color.sty' is not loaded}%
    \renewcommand\color[2][]{}%
  }%
  \providecommand\transparent[1]{%
    \errmessage{(Inkscape) Transparency is used (non-zero) for the text in Inkscape, but the package 'transparent.sty' is not loaded}%
    \renewcommand\transparent[1]{}%
  }%
  \providecommand\rotatebox[2]{#2}%
  \newcommand*\fsize{\dimexpr\f@size pt\relax}%
  \newcommand*\lineheight[1]{\fontsize{\fsize}{#1\fsize}\selectfont}%
  \ifx\svgwidth\undefined%
    \setlength{\unitlength}{428.16863457bp}%
    \ifx\svgscale\undefined%
      \relax%
    \else%
      \setlength{\unitlength}{\unitlength * \real{\svgscale}}%
    \fi%
  \else%
    \setlength{\unitlength}{\svgwidth}%
  \fi%
  \global\let\svgwidth\undefined%
  \global\let\svgscale\undefined%
  \makeatother%
  \begin{picture}(1,0.35624287)%
    \lineheight{1}%
    \setlength\tabcolsep{0pt}%
    \put(0.35223082,-0.01238127){\color[rgb]{0,0,0}\makebox(0,0)[lt]{\begin{minipage}{0.93369643\unitlength}\raggedright \end{minipage}}}%
    \put(0,0){\includegraphics[width=\unitlength,page=1]{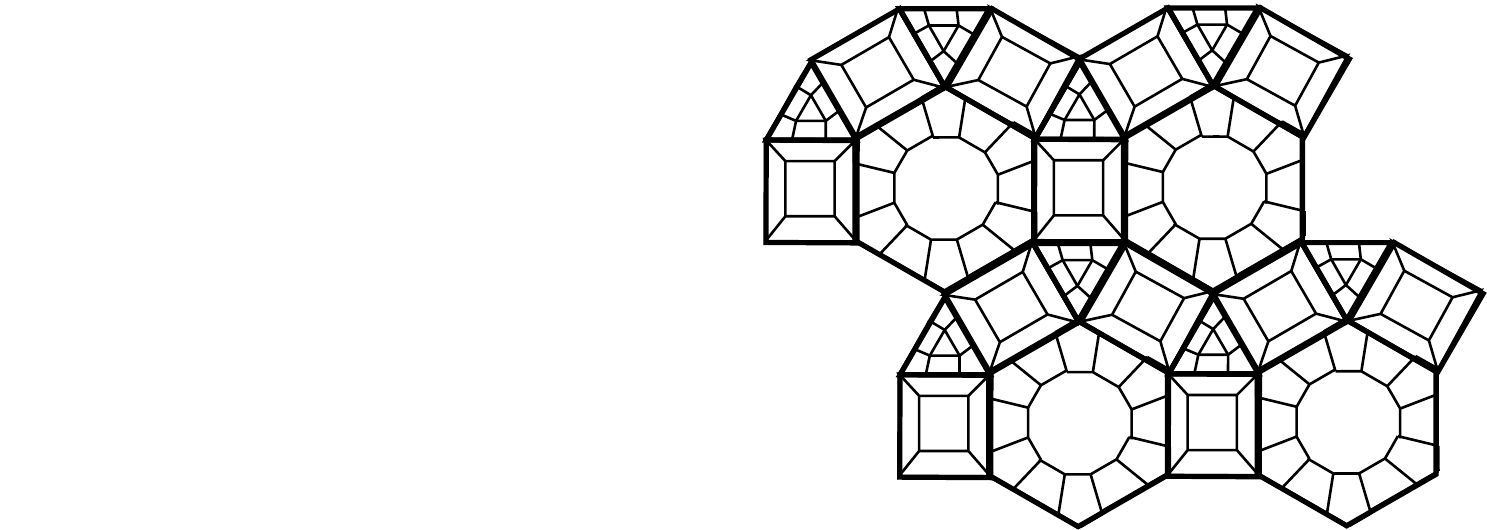}}%
    \put(0.01557595,0.09895684){\color[rgb]{0,0,0}\makebox(0,0)[lt]{\lineheight{1.25}\smash{\begin{tabular}[t]{l}(a)\end{tabular}}}}%
    \put(0,0){\includegraphics[width=\unitlength,page=2]{tLocusTile.pdf}}%
    \put(0.51951427,0.10912503){\color[rgb]{0,0,0}\makebox(0,0)[lt]{\lineheight{1.25}\smash{\begin{tabular}[t]{l}(b)\end{tabular}}}}%
  \end{picture}%
\endgroup%

\caption{Transfer Locus: (a) a pair  of data points is inserted on each edge between two simplicial points (b) Corresponding tiles  are peripheral subdivision of the former simplicial tile.  } \label{fig:t-locusTile} \end{figure} 

 \begin{figure}\centering   \def\svgwidth{\columnwidth}  
%% Accompanies image file 'vLocus.pdf' (pdf, eps, ps)
\begingroup%
  \makeatletter%
  \providecommand\color[2][]{%
    \errmessage{(Inkscape) Color is used for the text in Inkscape, but the package 'color.sty' is not loaded}%
    \renewcommand\color[2][]{}%
  }%
  \providecommand\transparent[1]{%
    \errmessage{(Inkscape) Transparency is used (non-zero) for the text in Inkscape, but the package 'transparent.sty' is not loaded}%
    \renewcommand\transparent[1]{}%
  }%
  \providecommand\rotatebox[2]{#2}%
  \newcommand*\fsize{\dimexpr\f@size pt\relax}%
  \newcommand*\lineheight[1]{\fontsize{\fsize}{#1\fsize}\selectfont}%
  \ifx\svgwidth\undefined%
    \setlength{\unitlength}{382.78248486bp}%
    \ifx\svgscale\undefined%
      \relax%
    \else%
      \setlength{\unitlength}{\unitlength * \real{\svgscale}}%
    \fi%
  \else%
    \setlength{\unitlength}{\svgwidth}%
  \fi%
  \global\let\svgwidth\undefined%
  \global\let\svgscale\undefined%
  \makeatother%
  \begin{picture}(1,0.18667521)%
    \lineheight{1}%
    \setlength\tabcolsep{0pt}%
    \put(0.66210475,-0.15737769){\color[rgb]{0,0,0}\makebox(0,0)[lt]{\begin{minipage}{1.04440391\unitlength}\raggedright \end{minipage}}}%
    \put(0.14478348,0.00413298){\color[rgb]{0,0,0}\makebox(0,0)[lt]{\lineheight{1.25}\smash{\begin{tabular}[t]{l}(a)\end{tabular}}}}%
    \put(0.41045156,0.0168707){\color[rgb]{0,0,0}\makebox(0,0)[lt]{\lineheight{1.25}\smash{\begin{tabular}[t]{l}(b)\end{tabular}}}}%
    \put(0.80790521,0.00495736){\color[rgb]{0,0,0}\makebox(0,0)[lt]{\lineheight{1.25}\smash{\begin{tabular}[t]{l}(c)\end{tabular}}}}%
    \put(0,0){\includegraphics[width=\unitlength,page=1]{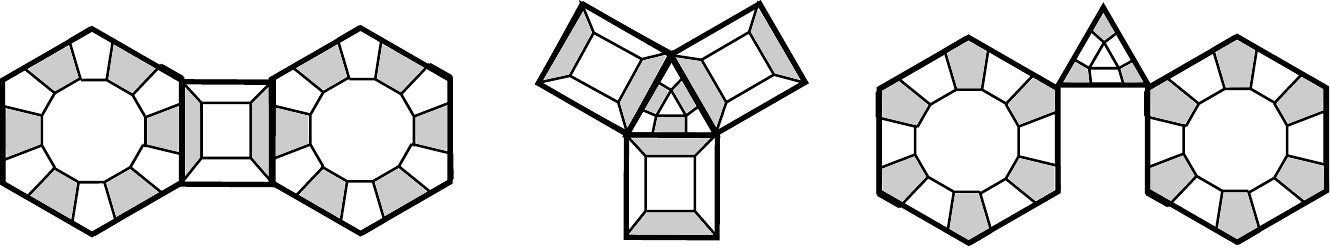}}%
  \end{picture}%
\endgroup%

\caption{  The six  transfer locus  grouped by pair of communicating locus (in gray) with adjacent pair of tiles: (a)   eV and vE tile, (b)   fE and eF  tile,  (c)  fV  and vF tile.} \label{fig:t-locus} \end{figure}
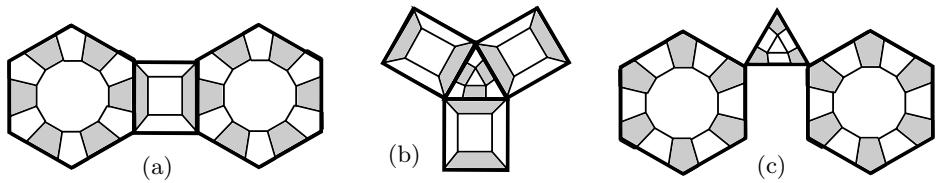 

%  \begin{figure}\centering   \def\svgwidth{\columnwidth}  
%   \includesvg{compositeReduction}
% \caption{ Basic operations: (a) Reductions and broadcast between an s-locus and its t-locus (b) Tranfer between two paired t-locus } \label{fig:basicoperations} \end{figure} 

\subsection{The transfer-locus}
From each of pair of simplicial locus $X,Y \in \{$V,E,F$\}$. We define two other locus (called ``{\it transfer locus}'') as follows: for each pair of adjacent points $p_X, p_Y$ in locus $X,Y$ we add two points $P_{yX}, P_{xY}$ dividing the segment $ [p_X, p_Y]$ in three. The six transfer locus defined in this way are called: eV, vE, eF, fE, vF, fV. The upper case designate the nearest simplicial locus, also called the father. The two transfer locus with identical father are called brother. 
Fig.~\ref{fig:t-locusTile} shows the tiling obtained by including the transfer locus within the seeds of the Vorono\"{i} Diagram.
The former simplicial Vorono\"{i} cell is subdivided: the peripheral regions represent the transfer locus, while the portion allocated to the simplicial locus is now reduced to   a central tile. 

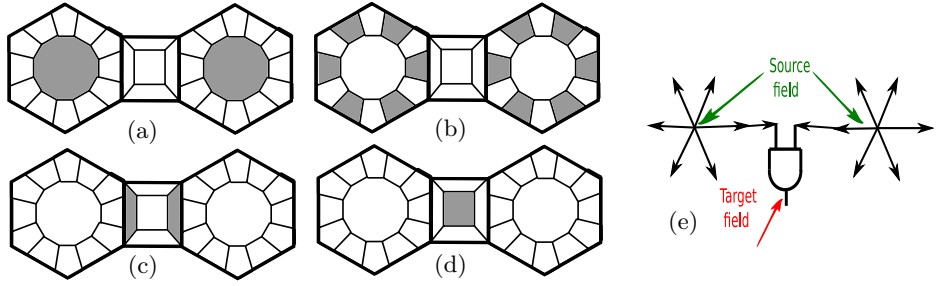
\begin{figure}\centering   \def\svgwidth{\columnwidth}  
%% Accompanies image file '3OpReduce.pdf' (pdf, eps, ps)
\begingroup%
  \makeatletter%
  \providecommand\color[2][]{%
    \errmessage{(Inkscape) Color is used for the text in Inkscape, but the package 'color.sty' is not loaded}%
    \renewcommand\color[2][]{}%
  }%
  \providecommand\transparent[1]{%
    \errmessage{(Inkscape) Transparency is used (non-zero) for the text in Inkscape, but the package 'transparent.sty' is not loaded}%
    \renewcommand\transparent[1]{}%
  }%
  \providecommand\rotatebox[2]{#2}%
  \newcommand*\fsize{\dimexpr\f@size pt\relax}%
  \newcommand*\lineheight[1]{\fontsize{\fsize}{#1\fsize}\selectfont}%
  \ifx\svgwidth\undefined%
    \setlength{\unitlength}{421.82416036bp}%
    \ifx\svgscale\undefined%
      \relax%
    \else%
      \setlength{\unitlength}{\unitlength * \real{\svgscale}}%
    \fi%
  \else%
    \setlength{\unitlength}{\svgwidth}%
  \fi%
  \global\let\svgwidth\undefined%
  \global\let\svgscale\undefined%
  \makeatother%
  \begin{picture}(1,0.30274787)%
    \lineheight{1}%
    \setlength\tabcolsep{0pt}%
    \put(0,0){\includegraphics[width=\unitlength,page=1]{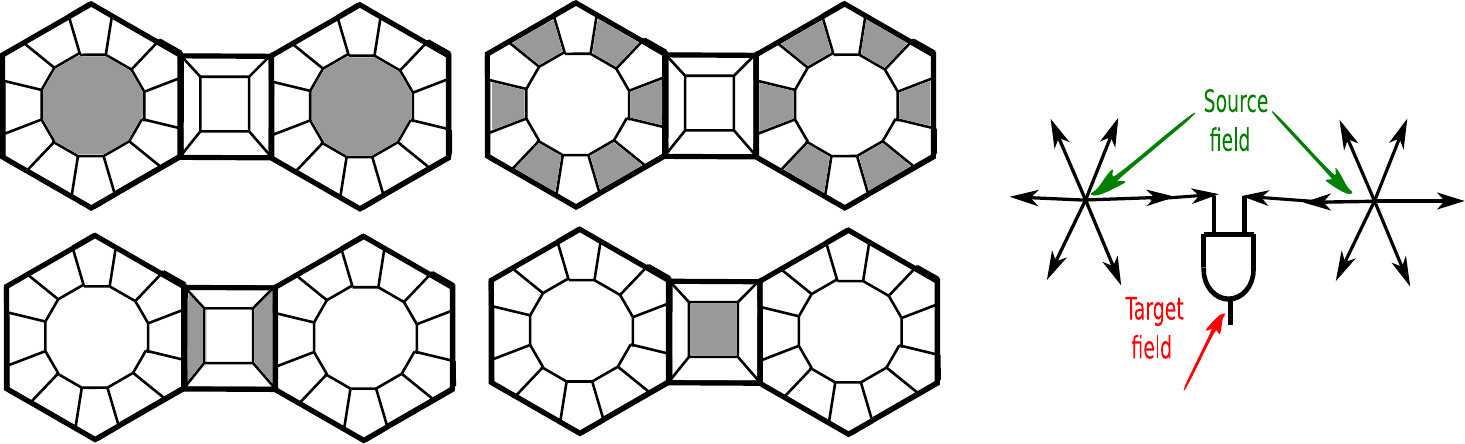}}%
    \put(0.12878855,0.15465114){\color[rgb]{0,0,0}\makebox(0,0)[lt]{\lineheight{1.25}\smash{\begin{tabular}[t]{l}(a)\end{tabular}}}}%
    \put(0.46031899,0.15701083){\color[rgb]{0,0,0}\makebox(0,0)[lt]{\lineheight{1.25}\smash{\begin{tabular}[t]{l}(b)\end{tabular}}}}%
    \put(0.12996839,0.00835305){\color[rgb]{0,0,0}\makebox(0,0)[lt]{\lineheight{1.25}\smash{\begin{tabular}[t]{l}(c)\end{tabular}}}}%
    \put(0.46031899,0.00953286){\color[rgb]{0,0,0}\makebox(0,0)[lt]{\lineheight{1.25}\smash{\begin{tabular}[t]{l}(d) \end{tabular}}}}%
    \put(0.71280118,0.05554601){\color[rgb]{0,0,0}\makebox(0,0)[lt]{\lineheight{1.25}\smash{\begin{tabular}[t]{l}(e)\end{tabular}}}}%
  \end{picture}%
\endgroup%

\caption{Decomposition of  $x\mapsto \forall^E(x)$. 
%The hexagons represents a vertex, and the two rectangles represents two edge.
(a) Initial boolV.
(b) Broadcast to a booleV
(c) Transfer to a boolvE 
(d) Reduction to a boolE
(e) Corresponding logical circuit.  
 } \label{fig:decomposition} \end{figure}

\paragraph{Decomposition of simplicial reductions in three steps.} Data traveling from a simplicial locus $X$ to another simplicial locus $Y$ will now transit through the     two intermediate   transfer  locus which form a pair of communicating data points, as shown in fig.~\ref{fig:t-locus}.
For example, bits  move  from V to E by passing through transfer locus eV, and then  vE,  inserted in-between the V and E locus. So the simplicial reduction $\forall^E$  is decomposed in three more elementary operations~(fig~\ref{fig:decomposition}:
 % Three operations are needed:
 \begin{enumerate} 
   \item operation $*^e$ broadcasts bits from each V-point to its 6 adjacent eV-points.
   \item operation $\uparrow$ transfers bits between the paired transfer locus eV and vE
   \item operation $/^\wedge$ computes the conjunction of bits on the two adjacent vE points.
 \end{enumerate}  
   
%\paragraph{Notation} 
The last step of pure reduction is noted using a slash and the reduction operation itself. 
So,   $\forall^E$  is now a function taking a boolV, producing a boolE, and programmed by composing three   operations:
\begin{equation}
 x\mapsto \forall^E(x)= /^\wedge(\uparrow(*^e(x))) = /^\wedge \uparrow *^e x. 
\end{equation}
In the second notation, we omit parenthesis for unary operations, this saves a lot of parenthesis.
%From a boolV, it computes a booleV then a boolvE they a boolE. 
 The superscript $e$ of broadcast $*^e$ reminds of the target locus.  
 Just like simplicial reductions, elementary operations are overloaded:    $*^e$ can    be applied to a boolV (resp. a boolF), it  broadcasts it to  a booleV (resp.   a booleF); Broadcast $*^v, *^f$ are defined similarly as $*^e$ .
 Transfer and reduction apply to any of the six transfer locus; 
 \paragraph{Compilation into a circuit.}
  A simplicial reduction corresponds to a  circuit part shown fig.~\ref{fig:decomposition}~(e) for $\forall^E$. Broadcast (resp. transfer, conjunction) is translated as a fan-out wiring, (resp. a "{\it trans-wire}"  crossing simplicial tiles , a logic gate). An operation-expression is compiled into a circuit, by putting together  the circuit parts associated to each of its operation. 
      %  For example $\forall^E$   cost 3*(2-1)=3 binary AND-gates.
%Elementary operations are written with symbols, such as  $*^e$, $\uparrow^E$, $\vee^E$.
%Their  role  is to complete the collection of  spatial-operations, by taking into account the particular circular ordering between the     vertices neighbor of a given vertex.

\subsection{Internal one-to-one communication between transfer locus}

\begin{figure} \centering %\def\svgwidth{\columnwidth}
 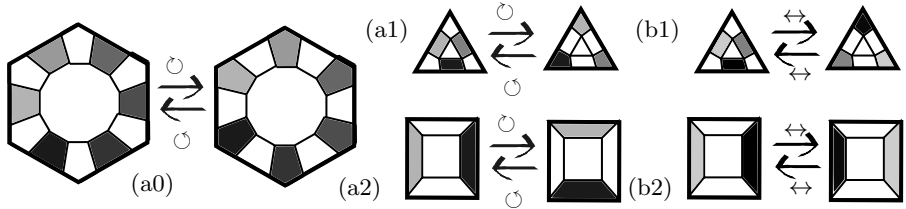
 \caption{ One-to-one communication between transfer locus, based on geometric transformation, applied on an integer field where different values are different gray tones.  (a0,a1,a2) clockwise/anticlockwise rotation $\circlearrowright, \circlearrowleft$   (b0,b1) central symmetry,$\leftrightarrow$  }
 \label{fig:operationT} \end{figure}
% 
% \begin{figure} \centering  \def\svgwidth{\columnwidth}  
%  \includesvg{centralSym}
%  \caption{ Central symmetry on t-locus  (a)$\leftrightarrow^E$: vE $\mapsto$ vE (b)$\leftrightarrow^F$: vF $\mapsto$ eF for triangulated planar graph  (c) composite communication  apex$^E$ : fV $\mapsto$ fE, wich uses  $\leftrightarrow^F$: vF $\mapsto$ eF .} 
%  \label{fig:centralSym} \end{figure}
%   

%\paragraph{operation on t-locus }   
     
  The   purpose of introducing transfer locus, is to increase the expressiveness of spatial types: 
  First of all, transfer fields such as  boolvE are used  often in the SDN-medium. A boolvE represent an edge together with an orientation: a vertex  can then compute wether it is in the inside or in the outside component.  
  Secondly, two new elementary communication-operations can be defined:     
 \begin{enumerate}              
   \item Clock, and anti-clock rotation map  each transfer locus to its brother.
   \item Central symmetry is available for Edge and Face transfer locus.
 %\item $\circlearrowright$  (resp $\circlearrowleft $ ):  eV $\mapsto $eV  clockwise (resp. conterclockwise) permutation.
  \end{enumerate}

  They 	are illustrated in fig.\ref{fig:operationT}~(a,b). 
We adopt the convention that one-to-one   communication are denoted with arrows.
Rotation (resp.  central symmetry) is noted $\circlearrowright$ (resp. $\leftrightarrow$). Transfer (already covered) was noted  $\uparrow$.
  Clock and anti-clock rotation can be defined because  two brother transfer locus  are interleaved. The central symmetry is an idempotent operation. It exploits the fact that for Edge and  Face,  the data-points of the two transfer locus are facing each other.  For faces, this is true  because face are always triangles. Note that this is the place where we use the hypothesis that  the planar-graph is maximal. The central symmetry maps an eF field to vF field and vice-versa. 
  For vertices, the central symmetry is not defined in the isotropic case, because as we have shown, the number of neighbors can very between 5,6 and 7.
 %Those graph are also called triangulated because all the faces are triangles, enabling the central symmetry between eF and vF points. 
  
  \paragraph{Six supplementary reductions.} By applying a reduction on the fields produced using the two opposite rotations, we can program a second set of six reductions mapping one transfer locus to its brother. For example,   when reducing with $\wedge$, we have a new function:
  \begin{equation}
  x\mapsto \mathrm{reduce2}^\wedge(x)=(\circlearrowright x)   \wedge (\circlearrowleft x) 
  \end{equation}

%  For vertices, the central symetry is not elementary, it can be obtained by composing three clockwise rotations.  
%    
% 
% \begin{figure} \centering  \def\svgwidth{\columnwidth}  
% \includesvg{compositeCom}
%  \caption{Composite communication (a) apex$^E$  (b)  Vertex$^V$.} 
%  \label{fig:compositeCom} \end{figure}
%   

\begin{figure} \centering  \def\svgwidth{\columnwidth}  
%% Accompanies image file 'apexComm.pdf' (pdf, eps, ps)
\begingroup%
  \makeatletter%
  \providecommand\color[2][]{%
    \errmessage{(Inkscape) Color is used for the text in Inkscape, but the package 'color.sty' is not loaded}%
    \renewcommand\color[2][]{}%
  }%
  \providecommand\transparent[1]{%
    \errmessage{(Inkscape) Transparency is used (non-zero) for the text in Inkscape, but the package 'transparent.sty' is not loaded}%
    \renewcommand\transparent[1]{}%
  }%
  \providecommand\rotatebox[2]{#2}%
  \newcommand*\fsize{\dimexpr\f@size pt\relax}%
  \newcommand*\lineheight[1]{\fontsize{\fsize}{#1\fsize}\selectfont}%
  \ifx\svgwidth\undefined%
    \setlength{\unitlength}{348.02926165bp}%
    \ifx\svgscale\undefined%
      \relax%
    \else%
      \setlength{\unitlength}{\unitlength * \real{\svgscale}}%
    \fi%
  \else%
    \setlength{\unitlength}{\svgwidth}%
  \fi%
  \global\let\svgwidth\undefined%
  \global\let\svgscale\undefined%
  \makeatother%
  \begin{picture}(1,0.26270208)%
    \lineheight{1}%
    \setlength\tabcolsep{0pt}%
    \put(0.21848959,-0.18449981){\color[rgb]{0,0,0}\makebox(0,0)[lt]{\begin{minipage}{1.14869515\unitlength}\raggedright \end{minipage}}}%
    \put(0,0){\includegraphics[width=\unitlength,page=1]{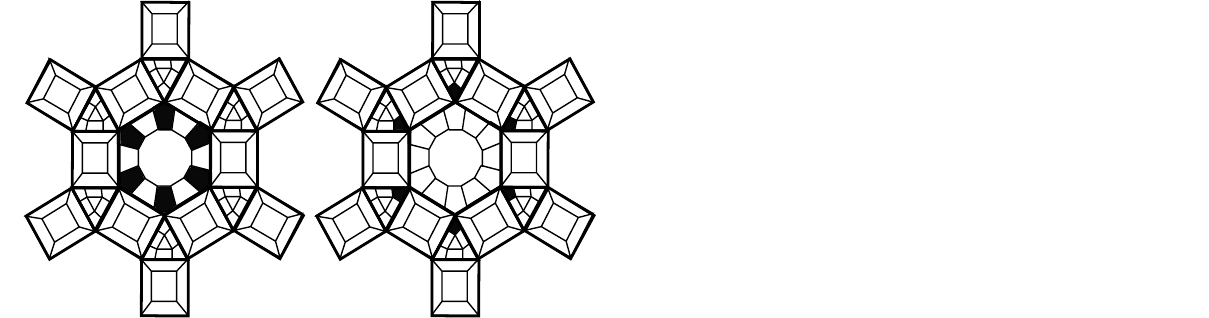}}%
    \put(-0.00296311,0.11545277){\color[rgb]{0,0,0}\makebox(0,0)[lt]{\lineheight{1.25}\smash{\begin{tabular}[t]{l}(a)\end{tabular}}}}%
    \put(0.24058588,0.12049698){\color[rgb]{0,0,0}\makebox(0,0)[lt]{\lineheight{1.25}\smash{\begin{tabular}[t]{l}(b)\end{tabular}}}}%
    \put(0,0){\includegraphics[width=\unitlength,page=2]{apexComm.pdf}}%
    \put(0.48872725,0.11672223){\color[rgb]{0,0,0}\makebox(0,0)[lt]{\lineheight{1.25}\smash{\begin{tabular}[t]{l}(c)\end{tabular}}}}%
    \put(0,0){\includegraphics[width=\unitlength,page=3]{apexComm.pdf}}%
    \put(0.73778144,0.12430539){\color[rgb]{0,0,0}\makebox(0,0)[lt]{\lineheight{1.25}\smash{\begin{tabular}[t]{l}(d)\end{tabular}}}}%
  \end{picture}%
\endgroup%

 \caption{ Apex communication.(a) fV field (b) transfer to a vF field (c)  central symmetry to a eF field(d)transfer to a fE field.} 
 \label{fig:apexComm} \end{figure}
  
\paragraph{Apex neighbors.}  \label{section:compcom} The central symmetry on faces is used to implement a composite communication called ``{\it apex}''.
 On a maximal planar graph, each edge  has    two distant vertices called  "{\it apex-vertices}", lying on the summit of the two triangles next to it.  Each vertex has also distant edges (5,6 or 7), also called apex-edges. 
The function $x\mapsto$apex$(x)$ implements a one-to-one composite communication from boolfV to boolfE, between a vertex and its apex-edges. The effect is illustrated in fig.~\ref{fig:apexComm}.  Bits transit from vertex to face, move within each face (central symmetry), and then from face to edge. Because of overloading, this function also implements the reciprocal transformation from edges to apex vertices:  apex$\circ$apex$=$Id, \footnote{and also a permutation between the eV-points of a vertex, and the  neighbor vertices}
 \begin{equation}
 x \mapsto \mathrm{apex}(x) =  \uparrow    \leftrightarrow     \uparrow x  
 \end{equation}

%   
% \begin{figure} \centering  \def\svgwidth{\columnwidth}  
% %\includesvg{VmeetingPoint} 
% \caption{ Vmeet-points and E-meet-points  of  2 $x$-blobs in gray, beyond the border is considered false 
% %(b) the field nbcc$(x)$ 
% (a) a merge-vertex and a merge-edge (b) a div-vertex and three div-edge.} 
%  \label{fig:VmeetingPoint} \end{figure}
%   
% 
% \begin{figure} \centering  \def\svgwidth{\columnwidth}  
% %\includesvg{VmeetingPoint} 
% \caption{ Detecting V Meet-points  of  2 $x$-blobs in gray as vertice where the local number of connected components, nbcc$(x)$ is $\geq$ 2. (a)ringChange$(x)=$ apex$^V$(frontier$^E(x)$)  (b)   $\mathrm{nbcc}(x) =   (+^V(\mathrm{ringChange}(x))/ 2$.} 
%  \label{fig:VmeetingDetection} \end{figure}
%  
 
% 
% \begin{figure} \centering  \def\svgwidth{\columnwidth}  
% %\includesvg{VmeetingPoint} 
% \caption{ Detecting E Meet-points  of  2 $x$-blobs in gray as edge
% (a) Faces outside the frontier: $\forall^F(\neg\mathrm{frontier}^E(x))$
% (b) Edge at the center of rhombus outside the frontier: $\forall^\diamond(\neg\mathrm{frontier}^E(x))= \forall^E(\forall^F(\neg\mathrm{frontier}^E(x)))$.
% (c) $\mathrm{frontier}^V(x)$ 
% (d) Inside of frontier$^V(x): \forall^E(\mathrm{frontier}^V(x) ) $
% edges and normal edges of $\mathrm{frontier}^E(x))$ are marked in gray, the other edges are in black, they  are edges adjacent to locally distinct frontier.
% } 
%  \label{fig:EmeetingDetection} \end{figure}

\begin{figure} \centering  \def\svgwidth{\columnwidth}  
%% Accompanies image file 'VmeetingPoint.pdf' (pdf, eps, ps)
\begingroup%
  \makeatletter%
  \providecommand\color[2][]{%
    \errmessage{(Inkscape) Color is used for the text in Inkscape, but the package 'color.sty' is not loaded}%
    \renewcommand\color[2][]{}%
  }%
  \providecommand\transparent[1]{%
    \errmessage{(Inkscape) Transparency is used (non-zero) for the text in Inkscape, but the package 'transparent.sty' is not loaded}%
    \renewcommand\transparent[1]{}%
  }%
  \providecommand\rotatebox[2]{#2}%
  \newcommand*\fsize{\dimexpr\f@size pt\relax}%
  \newcommand*\lineheight[1]{\fontsize{\fsize}{#1\fsize}\selectfont}%
  \ifx\svgwidth\undefined%
    \setlength{\unitlength}{468.07229054bp}%
    \ifx\svgscale\undefined%
      \relax%
    \else%
      \setlength{\unitlength}{\unitlength * \real{\svgscale}}%
    \fi%
  \else%
    \setlength{\unitlength}{\svgwidth}%
  \fi%
  \global\let\svgwidth\undefined%
  \global\let\svgscale\undefined%
  \makeatother%
  \begin{picture}(1,0.18778873)%
    \lineheight{1}%
    \setlength\tabcolsep{0pt}%
    \put(-0.00189336,0.07447213){\color[rgb]{0,0,0}\makebox(0,0)[lt]{\lineheight{0}\smash{\begin{tabular}[t]{l}(a)\end{tabular}}}}%
    \put(0.33477369,0.08913014){\color[rgb]{0,0,0}\makebox(0,0)[lt]{\lineheight{0}\smash{\begin{tabular}[t]{l}(b)\end{tabular}}}}%
    \put(0.67929567,0.09489742){\color[rgb]{0,0,0}\makebox(0,0)[lt]{\lineheight{0}\smash{\begin{tabular}[t]{l}(c)\end{tabular}}}}%
    \put(0,0){\includegraphics[width=\unitlength,page=1]{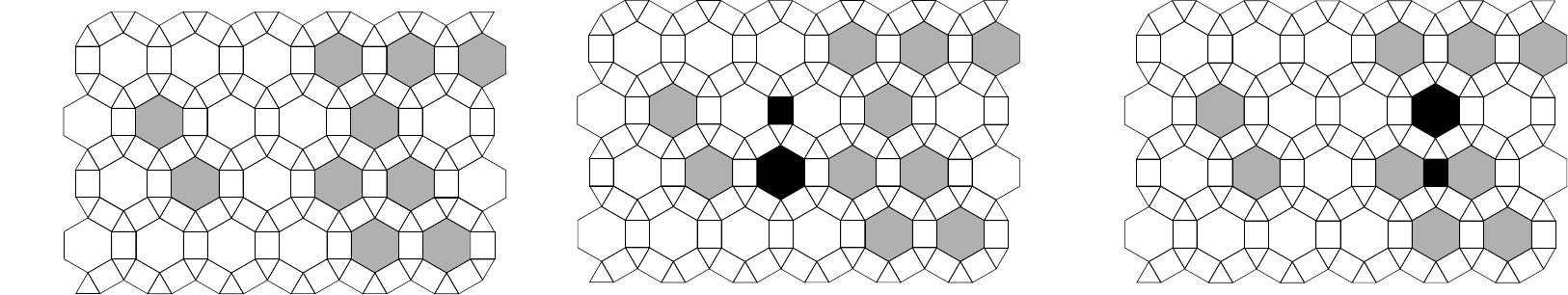}}%
  \end{picture}%
\endgroup%
 
\caption{ Meet-points in black,   (a) 2 $x$-blobs in gray, $x$=0 beyond the border.
%(b) the field nbcc$(x)$ 
(b) a merge-vertex and a merge-edge (c) a div-vertex and a div-edge.} 
 \label{fig:VmeetingPoint} \end{figure}
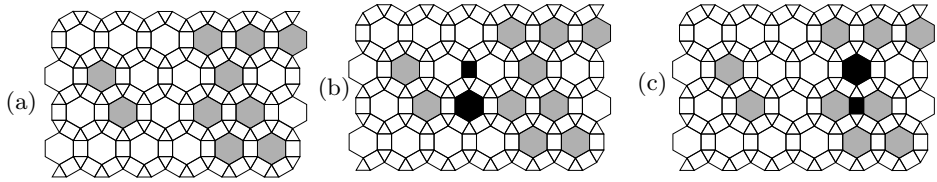 
  
\section{Computing  the meet-point function  $x\mapsto$ meet$(x)$ } \label{sec:meet}
\paragraph{The vertex frontier, its inside and outside.}
Let $x$ be a boolV representing. Before defining meet-points $x\mapsto$ meet$(x)$ , we first need to
compute the set of vertices adjacent to the frontier of $x$-blobs: 
\begin{equation}
x \mapsto \mathrm{frontier}^V(x)=\exists^V(\mathrm{frontier}^E(x)),
\end{equation}
 Frontier$^V(x)$ is decomposed into an inside frontier 
$\mathrm{frontier}_\mathrm{in}^V(x)=x\wedge\mathrm{frontier}^V(x)$, and an outside frontier: 
$\mathrm{frontier}_\mathrm{out}^V(x)=\neg x\wedge\mathrm{frontier}^V(x).$

\paragraph{Global meet-points.} The $x$ blobs represent  agent's supports. For illustrating cellular circuits, we program the function    $x\mapsto$ meet$(x)$  used in SDN-media  for preserving the  supports when agents move. We will use it in section~\ref{subsec:VD} to compute the discrete Vorono\"{i} diagram. 
It has a boolV and a boolE component: 
meet$(x)=($meet$^V(x),$meet$^E(x))$.
Each of those component is a conjunction of a div and a merge part:
 meet$^V(x)=$merge$^V(x)\vee$div$^V(x)$ and meet$^E(x)=$merge$^E(x)\vee$div$^E(x)$. 

\begin{definition}\label{def:meeting} 
Let $x$ be a bool$V$, merge$^V(x)$ is true for empty vertices (resp. merge$^E(x)$ is true for edges)  adjacent to two (resp. the out-frontier$^V$ of two) distinct  $x$-blobs.   div$^V(x)$ is true for filled vertices, (resp. div$^E(x)$ true for edges),  adjacent to two distinct $x$-holes (resp. the in-frontier$^V$ of two  distinct  $x$-holes). 
 \end{definition} 
  
\paragraph{Meet-points are needed to preserve connectedness.}  In a SDN-medium,  agents move by modifying   their $x$-blob support, by emptying (resp. filling) a given vertex  on the inside (resp. outside) frontier. However, such a move  can cause a division (resp. a merge)
%Detection of meet-points is necessary for preserving $x$-blobs.
if the chosen vertex is a  div-vertex  (resp. a merge-vertex).
The same hold if two vertices adjacent to a div-edge (resp. to a merge-edge) are simultaneously emptied (resp. filled). In summary, preserving $x$-blob's supports implies not modifying  meet-vertices, and not modifying simultaneously the two vertices on both sides of a meet-edge (see fig.~\ref{fig:VmeetingPoint}).

\paragraph{Local  meet-points.}  An $x$-blob can be arbitrary big, and with non-convex shape. Computing whether two vertices belong to the same $x$-blob requires the exploration of a region which is not a priori bounded. It cannot be done with a fixed operation-expression wich can explores only a fixed radius neighborhood. We propose an alternative definition of meet-points: local meet-points.  In definition~\ref{def:meeting}, instead of $x$-blobs, we take the  $x$-blobs locally induced in the immediate neighborhood, by intersecting with a ball of a given radius $r$, centered on the   meet-point. For meet-vertices (resp. meet-edges) we use $r=2$ (resp. $r=3$).  Local meet-points are not necessary global. This is because although locally one may find two distinct components, those two components may meet if we look further.
  What matters is that global meet-points are also local meet-points, so detecting local meet-points is an overkill but work for our purpose of preserving $x$-blobs.
 
 \subsection{Computing the Meet-vertex function,  $x\mapsto$ meet$^V(x)$.} 
\begin{figure} \centering \def\svgwidth{\columnwidth}  
  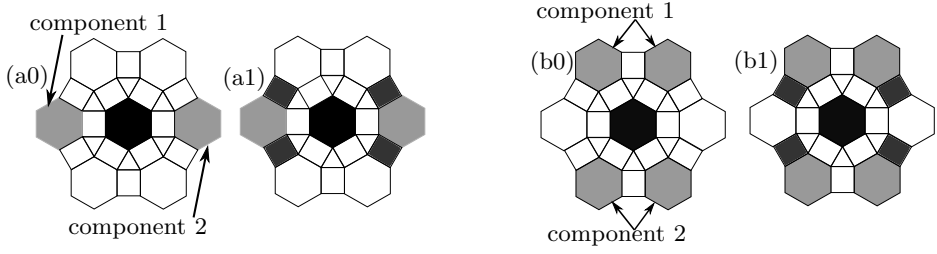 
\caption{Detecting (a) the merge-vertex, and (b) the dividing-vertex of fig.~\ref{fig:VmeetingPoint}~(b,c). The added darker rectangles in (a1,b1) represent the edge in frontier$^E(x)$, which are also apex  edges of the central vertex. There are four of them in both case.  
%For meeting-vertices, we color one component of the ring in black and the other one in light gray.
 } \label{fig:detectMeetingVV} \end{figure}

\paragraph{Computing  $x\mapsto$ nbcc$(x)$.} 
As shown in fig.~\ref{fig:detectMeetingVV},  the radius-2 ball centered on a vertex includes a ring of 6 neighbor vertices\footnote{It could be 5 or 7 for the isotropic case}; 
 As a prerequisite, we need to compute  the number nbcc$(x)$ of filled connected components in this ring.
  For example, in fig.~\ref{fig:detectMeetingVV}~(a) and (b) nbcc$(x)=2$. In general the number of neighbors is $\leq7$, hence nbcc$(x)\leq3$ so nbcc$(x)$ is an intV2 (encoded with 2 bits). 
  % We use  frontier$^E(x)$ shown in fig.\ref{fig:feature}~(c).
   Each  component is delimited by two apex-edges  in frontier$^E(x)$,  we therefore only need to   make the sum of those and divide by two. 
   \begin{equation}
     \mathrm{nbcc}(x) =   (/^+ (\mathrm{apex}(\mathrm{frontier}^E(x))))/2 .
   \end{equation}

\paragraph{Computing Meet-vertice.}
If nbcc$(x)=2$, filling (resp. emptying)   merges   those 2 (divide into those 2) components. If nbcc$(x)=3$  the same reasoning applies with 3 components. Finally, if nbcc$(x)\leq 1$, no division nor merge happens, hence:

%$\#_\mathrm{gate}($meet$^V(x)$) = 23+1=24 .
 
\begin{equation}
\mathrm{meet}^V(x)=\mathrm{nbcc}(x) \geq 2,~~\mathrm{div}^V(x)=\mathrm{meet}^V(x)\wedge  x,~~\mathrm{merge}^V(x)=\mathrm{meet}^V(x)\wedge\neg x
\end{equation}

Computing frontier$^E(x)$ costs 3 gates. nbcc$(x) \geq 2  \Leftrightarrow /^+ (\mathrm{apex}(\mathrm{frontier}^E(x)))\geq4$; Knowing that $x$ is even, the computation $x\mapsto/^+(x)\geq 4$  can be done with only 17 gates. 
Finally, meet$^V$ costs $17+3=20$ gates, div$^V$ costs  21 gates, and merge$^V$ costs  22 gates. %20 au lieu de 23 corriger
 
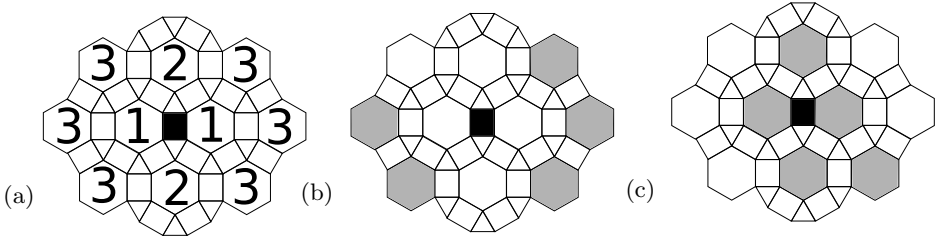
\begin{figure} \centering \def\svgwidth{\columnwidth}  
%% Accompanies image file 'detectMeetingEE.pdf' (pdf, eps, ps)
\begingroup%
  \makeatletter%
  \providecommand\color[2][]{%
    \errmessage{(Inkscape) Color is used for the text in Inkscape, but the package 'color.sty' is not loaded}%
    \renewcommand\color[2][]{}%
  }%
  \providecommand\transparent[1]{%
    \errmessage{(Inkscape) Transparency is used (non-zero) for the text in Inkscape, but the package 'transparent.sty' is not loaded}%
    \renewcommand\transparent[1]{}%
  }%
  \providecommand\rotatebox[2]{#2}%
  \newcommand*\fsize{\dimexpr\f@size pt\relax}%
  \newcommand*\lineheight[1]{\fontsize{\fsize}{#1\fsize}\selectfont}%
  \ifx\svgwidth\undefined%
    \setlength{\unitlength}{168.19849243bp}%
    \ifx\svgscale\undefined%
      \relax%
    \else%
      \setlength{\unitlength}{\unitlength * \real{\svgscale}}%
    \fi%
  \else%
    \setlength{\unitlength}{\svgwidth}%
  \fi%
  \global\let\svgwidth\undefined%
  \global\let\svgscale\undefined%
  \makeatother%
  \begin{picture}(1,0.25347595)%
    \lineheight{1}%
    \setlength\tabcolsep{0pt}%
    \put(0.31614166,0.03890491){\color[rgb]{0,0,0}\makebox(0,0)[lt]{\lineheight{0}\smash{\begin{tabular}[t]{l}(b)\end{tabular}}}}%
    \put(-0.00438759,0.03553609){\color[rgb]{0,0,0}\makebox(0,0)[lt]{\lineheight{0}\smash{\begin{tabular}[t]{l}(a)\end{tabular}}}}%
    \put(0,0){\includegraphics[width=\unitlength,page=1]{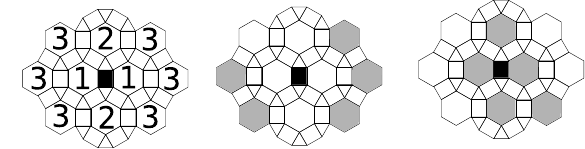}}%
    \put(0.66656128,0.04221383){\color[rgb]{0,0,0}\makebox(0,0)[lt]{\lineheight{0}\smash{\begin{tabular}[t]{l}(c)\end{tabular}}}}%
  \end{picture}%
\endgroup%
 
\caption{Detecting the two meet-edges of  figure~\ref{fig:VmeetingPoint},(a) the radius 3 ball centered on an edge contains vertices at distance 1,2 and 3   (b) merging edge  (c) dividing edge. 
%For meeting-vertices, we color one component of the ring in black and the other one in light gray.
 } \label{fig:detectMeetingEE} \end{figure}

\subsection{Computing the meet-edge function  $x\mapsto$meet$^E(x)$.}   In order to compute a local version of meet$^E(x)$ for one edge, we consider the radius 3 ball centered on that edge, shown  fig.~\ref{fig:detectMeetingEE}.
 It includes  three kinds of neighbor vertices:  two immediate neighbors at distance 1, two apex neighbors at distance 2, and 6 remote neighbors at distance 3.  Immediate and apex neighbors form a rhombus.
 We will need a function $x\mapsto \forall^\diamond(x)$
 which takes a boolV $x$  and computes a boolE  true for an edge if $x$ is true  within the rhombus centered on that edge. It can  be computed using $3+4=7$ gates by chaining two   reductions. The formula can also be applied to a boolE. 
 \begin{equation}
  x \mapsto \forall^\diamond(x)= \forall^E(\forall^F(x)) 
 \end{equation}
 An edge $y$ is locally merging two $x$-blobs (  fig.~\ref{fig:detectMeetingEE}~(b)) if there are two locally induced $x$-blobs, two vertices away from each other, one vertex away on each side of the edges. Equivalently:
(i) the surrounding rhombus is empty (otherwise those two blobs would meet locally)
(ii)  on each side of the rhombus, some vertices at distance 3, must be full. (ii) is  checked if and only if  both immediate neighbor of $y$  belong to Frontier$^V(x)$, this  is computed as $\forall^E(\mathrm{Frontier}^V(x))$. Putting  together (i) and (ii) we obtain:

\begin{equation}    
\mathrm{merge}^E(x)=\forall^\diamond(\neg x)\wedge\forall^E(\mathrm{frontier}^V(x))\end{equation} 

Because  a div-edge is a merge-edge of the complement we obtain: 
 \begin{equation} \mathrm{div}^E(x)=\forall^\diamond(x)\wedge\forall^E(\mathrm{frontier}^V(\neg x))  
\end{equation}  
 
But $\mathrm{frontier}^V(\neg x)=\mathrm{frontier}^V(x)$. From
 $(\forall^\diamond(x) \vee  \forall^\diamond(\neg x))=\forall^\diamond(\neg\mathrm{frontier}^E(x))$ we can factorize, simplify  and derive:
  
 \begin{equation}\label{eq:meetE} 
\mathrm{meet}^E(x)=  \forall^\diamond(\neg\mathrm{frontier}^E(x)) \wedge\forall^E(\mathrm{frontier}^V(x))  
\end{equation} 
 
The auxiliary field Frontier$^E(x)$ has already been computed  for nbcc$(x)$, so it is availabe and free! The field frontier$^V(x) = \exists^V$(frontier$^E(x))$ costs 5 gates.   $x\mapsto\forall^\diamond( x)$  costs   7 gates, the non spatial conjunction $\wedge$ applied to a boolE costs 3 gates. meet$^E(x)$ costs  7 + 3 + 5 + 3 = 18 gates. The radius is 3.  Taken separately, div$^E(x)$ costs 7 + 3 + 3 + 5 = 18 gates. merge$^E(x)$ cost 19 gates.  If meet$^E$ has aleady been computed, then one can  also compute div$^E(x)$=meet$^E\wedge$ inside$^E(x)$ and  merge$^E(x)$= meet$^E\wedge$ inside$^E(\neg x)$, which adds only 6 gates. 

\section{Sequential  cellular circuits}
 
 A set of fields is called a configuration. A sequential circuit is described  by a function %programmed using spatial operation,
   updating a configuration, i.e. with identical domain and co-domain. % We call it a ``cellular sequential circuit''.
  
% \paragraph{Iterating operation-expression.} We compute a function $f$ such that the domain and codomain are the same, and $f$ can be iterated, and models a seqential circuit. 
  
 \begin{definition}Let $\tau_{i=1\ldots n}$ be some spatial-types and  $\Gamma= \prod_{i=1}^n \tau_i$.
 A sequential Cellular Circuit  is  a mapping $f: \Gamma\mapsto \Gamma$, each component  $f_i$  is computed as an operation-expression \end{definition} 
 %The word ``Cellular'' conveys the idea of translation-invariance implied by the underlying spatial-type. 
 We often call it simply a ``cellular circuit''.
  Starting from the initial configuration $x^0=(x^0_0\ldots x^0_k)$, we iterate $t$ times and obtain the configuration at time $t$: $x^t=f^t(x^0)= f(f^{t-1}(x^0)) = (x^t_0\ldots x^t_k)$.   The  sequence  $x^0, x^1,\ldots, x^t$    represents the  circuit iteration.   A   component $(x^t_i)_{t\in \mathbb{N}}$  represent the successive values (of type $\tau_i$) of a stored field, which  is called a ``{\it layer}''.
  % If $l$ is a layer's name, we call the updated value   $l_\mathrm{new}$.
 % The layers represent information which is stored, in-between two  iterations. 
 
 \paragraph{Circuit tiling} The set of gates can be partitioned with one vertex per tile,  by choosing a vertex responsible for each edges and faces. In the hexagonal case, this can be done in a simple systematic way, by selecting direction,  and in particular, each class correspond to an identical circuit tile. Fig.~\ref{fig:t-locusTile} shows 4 such tiles.   In the isotropic case,     a distributed algorithm must decide using random tournament and makes sure that each PE get approximately the same number of edges and faces.
 
 \paragraph{Complexity of a  cellular circuit} It is measured by \begin{enumerate}
   \item the radius  and gate density of the updating function,
   \item the memory density equal to the sum of the bit density of layers.
   \item the trans-wire count which is the   number of wires crossing a circuit tile.
 \end{enumerate}
  In this introductory paper,  we consider  only simple circuits with a single boolV layer. Therefore they have a minimal memory density of 1.  
 
\begin{figure} \centering \def\svgwidth{\columnwidth} 
   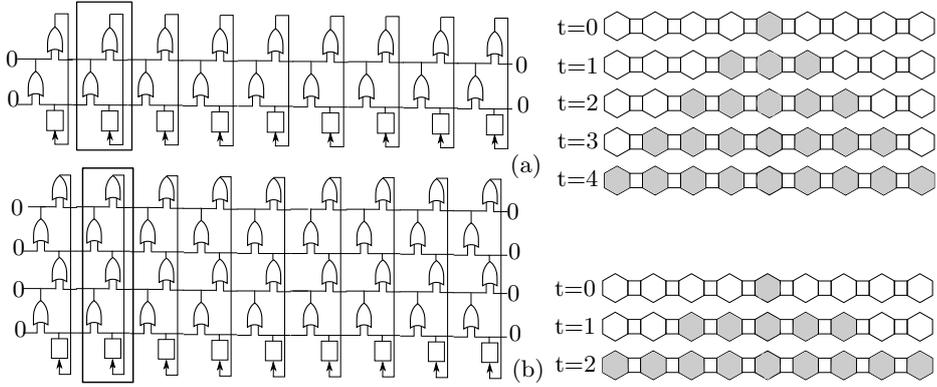
\caption{The 9-vertice 1D-circuit  obtained from: (a)  $x\mapsto$ neighborhood$^V(x)$   (b)  $x\mapsto$ neighborhood$^V$(neighborhood$^V(x)$). The rectangle delimits a tile. Squares are 1-bit register storing $x$. Execution is illustrated with an initial configuration having  a single central true vertex.  }
\label{fig:growCircuit} \end{figure}

\subsection{Analysis of a simple cellular circuit for growing blobs.}\label{subsection:complexity}
   The  boolV circuit  $x\mapsto$ neighborhood$^V(x)=\exists^V(\exists^E(x))$   let some initial $x^0$-blobs grow, until they meet and merge, and fill the whole medium. Its gate density is 8. This simple circuit is helpfull to better understand the different concepts involved. 
   %Radius is 2, gate-count is 8.
  Consider   a 1D   graph  consisting of a simple   line of 9 vertices.   For this ``degenerated'' planar graph, the gate density  is 2 instead of 8. We can easily represent the compiled circuit in fig.~\ref{fig:growCircuit}~(a).
  It is made of 9 copies of the same tile. 
 
   In fig.~\ref{fig:growCircuit} we can see that each PEs computes the boolE $\exists^E(x)$ of its left edge, in the lower row of or-gate. The upper row computes the $\exists^V$.
Zero values (the neutral value of OR) must be supplied to the OR-gate, to the border tiles. 

  The circuit for  $x\mapsto$   neighborhood$(x)$ has gate density 2, radius   2,   and   trans-wire count also 2. The circuit  $x\mapsto$ neighborhood( neighborhood$(x)$ )  shown in  fig.~\ref{fig:growCircuit}~(b) goes two times faster.  Gate density, radius and  trans-wire count  are all doubled to  4. More generally,   $x\mapsto$ neighborhood$^k(x))$, $k>2$,  compiles into a cellular-circuit with a gate density,radius and trans-wire count of $2*k$,  and   goes $k$ times faster. 
 Translated in a   CA framework, each PE would need to explore a neighborhood of radius $k$ which is $O(k^2$) for the 2D homogeneous planar graph. 
 This circuit family is very particular because the same reduction (OR) is applied repetitively $2k$ times. In the general case,   different reductions are applied at each stage. If the radius is $r$, for each stage at height $h$, $1\leq h\leq r$, a CA simulation must entirely traverse the neighborhood of radius $r-h$. The complexity of the translated CA execution becomes $O(\Sigma_{h=1}^r (r-h)^2 )=O(r^3$) instead of $O(r)$ for cellular circuits. The improved complexity is due to a fine grain interleaving of computation with communication: As soon as a field is computed by applying a reduction, it is communicated again. In this way, a reduction done for one vertex benefits to neighbor vertices.
 As a result, the  circuit's  complexity augments  only linearly with the radius. When programming complex circuits, this feature allows to handle  large radius.
  
%The depth of the circuit also augments linearly. 
   
%    
% \begin{figure} \centering  \def\svgwidth{\columnwidth}   
% \includesvg{VDwave}$
% \caption{Iteration of the circuit computing the strict Vorono\"{i} cells (light gray) at $t_c=3$. At $t=0$, $x^0$ is the seeds. The VD itself progressively appears as merge$(x^t)$, in black, plus the unique outside$^V(x^3)$ tri-vertex (in dark gray)}
% \label{fig:VD} \end{figure}  
%      

 \subsection{A circuit for the discrete Vorono\"{i} Diagram (VD) of $x$-blobs.} \label{subsec:VD}
\label{subsection:VD}
 
\paragraph{A definition of discrete VD encoded with fields.}
Let $x^0$ be a boolV encoding  seeds as  $x^0$-blob. The seeds are therefore possibly non-punctual.  The discrete Vorono\"{i} cell of an $x^0$-blob is the set of vertices strictly nearer to it than to other blobs.
Here, the distance is the same  V,E,F hop count, used to define the radius.   Vorono\"{i} cells partition the set of vertices.
The VD is usually defined as the partition itself. 
 In the continuous case, this partition can be represented by polygons; In the discrete case it is less obvious due to the following discrete artifact:
If two nearest seeds are at even (resp. odd) distance, their Vorono\"{i} cell are separated by an edge (resp. a vertex).  
Spatial types can   represent    a set of vertices (boolV) as well as a set of edges (boolE). This allows to define the VD as a boolE and a boolV: VD($x$)=(VD$^V(x)$, VD$^E(x)$):

\begin{definition}
The  discrete Vorono\"{i} Diagram (VD) of a set of $x$-blobs is the set of edges and vertices equidistant to at least two nearest  $x$-blobs. 
\end{definition}
Let closure$^V: $boolV$\times$boolE$\mapsto$boolV be defined as closure$^V(x,y)=x\vee\exists^V(y)$. It needs 4 gates. We remark that closure$^V($VD$(x))$ encodes VD$(x)$ using only vertices, with the  additional important topological property of separating the seeds. This property is of high interest for SDN-media, and makes it a legitimate representation of the VD. We call it the ``{\it vertex-VD}''.

\begin{figure} \centering \def\svgwidth{11cm  }   
 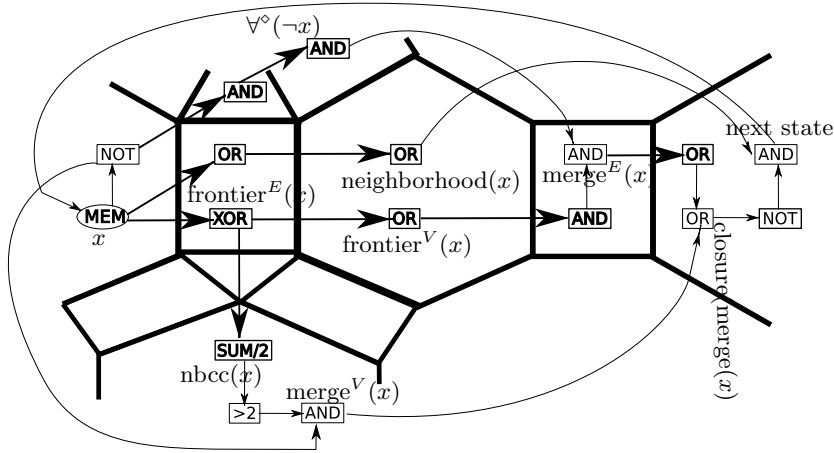
\caption{Folded form of the VD circuit. Thick arrows represent multiple  connections to all neighbors: the fan out (resp fan in) si the co-arity of the sender (resp. the receiver). Thick gate represent simplicial reduction, they combine a number of inputs equal to the co-arity. Curved arrows represent a single connection, it links two gates of distinct radius. Thin arrows (including curved) and thin gates represent non-spatial computation within the same tile. To unfold the circuit, thick arrow must be drawn towards every neighbors, and each gate must be copied in each tile. }
\label{fig:circuitVoronoi} \end{figure}

\begin{figure} \centering  %\def\svgwidth{\columnwidth  }   
 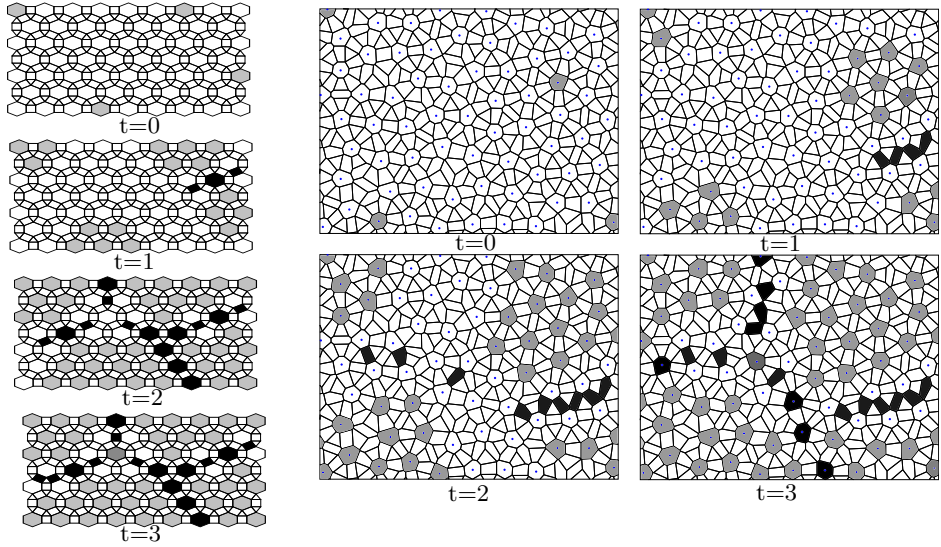
\caption{Iteration of the circuit computing the strict Vorono\"{i} cells  in a boolV layer $x^t$ (light gray). At $t=0$,   $x^0$ contains  the seeds.  Convergence happens at $t_c=3$.   Two  auxiliary boolV and boolE field (black) represent merge$(x^t)$. There is also a unique multi-vertex (dark gray) detected as outside$^V(x^3)$. 
Execution on the left uses the hexagonal medium, and on the right, the isotropic medium of fig.~\ref{fig:VEFtiling}.}
\label{fig:VD} \end{figure}

\paragraph{The VD circuit.} Instead of marking the vertex-VD, we mark the complement which consists  in Vorono\"{i} cells  deprived from  vertices  adjacent to  another distinct Vorono\"{i} cell. We  call those ``{\it strict Vorono\"{i} Cell}''.
We  reuse the preceding growing circuit of the preceding subsection. Starting from $x^0$-blobs, we grow  everywhere except on closure$^V($merge$(x))$ where merge$(x)$ = (merge$^V(x)$, merge$^E(x)$); 
Since  merge-points were defined precisely so as to avoid merging supports, the growth will be canceled   when two $x^t$-blobs come close (one or two vertices away). As a result, the $x^t$-blobs  will grow until they exactly fill their associated strict Vorono\"{i} cell.
As shown  in fig.~\ref{fig:VD},  convergence happens in a time $t_c$ equals to  half the diameter of the medium. We choose a set of seeds in order to  illustrate  a ``multi-vertex''  equidistant to three seeds or more. In the hexagonal case, multivertice can also be detected because they are in outside$^V(x^{t_c})$, i.e. they remain surrounded by unmarked vertices.  
\begin{theorem}
The circuit $x\mapsto$neighborhood$(x)\wedge\neg$closure$^V($merge$(x))$ fills exactly all the strict Vorono\"{i} cell, using 55 gates and a radius 4.
% and leaves empty the closure$^V$ of the Vorono\"{i} diagram.
\end{theorem}

\begin{figure} \centering  \def\svgwidth{\columnwidth  }   
%% Accompanies image file 'triVertex.pdf' (pdf, eps, ps)
\begingroup%
  \makeatletter%
  \providecommand\color[2][]{%
    \errmessage{(Inkscape) Color is used for the text in Inkscape, but the package 'color.sty' is not loaded}%
    \renewcommand\color[2][]{}%
  }%
  \providecommand\transparent[1]{%
    \errmessage{(Inkscape) Transparency is used (non-zero) for the text in Inkscape, but the package 'transparent.sty' is not loaded}%
    \renewcommand\transparent[1]{}%
  }%
  \providecommand\rotatebox[2]{#2}%
  \newcommand*\fsize{\dimexpr\f@size pt\relax}%
  \newcommand*\lineheight[1]{\fontsize{\fsize}{#1\fsize}\selectfont}%
  \ifx\svgwidth\undefined%
    \setlength{\unitlength}{349.04711049bp}%
    \ifx\svgscale\undefined%
      \relax%
    \else%
      \setlength{\unitlength}{\unitlength * \real{\svgscale}}%
    \fi%
  \else%
    \setlength{\unitlength}{\svgwidth}%
  \fi%
  \global\let\svgwidth\undefined%
  \global\let\svgscale\undefined%
  \makeatother%
  \begin{picture}(1,0.27778676)%
    \lineheight{1}%
    \setlength\tabcolsep{0pt}%
    \put(0,0){\includegraphics[width=\unitlength,page=1]{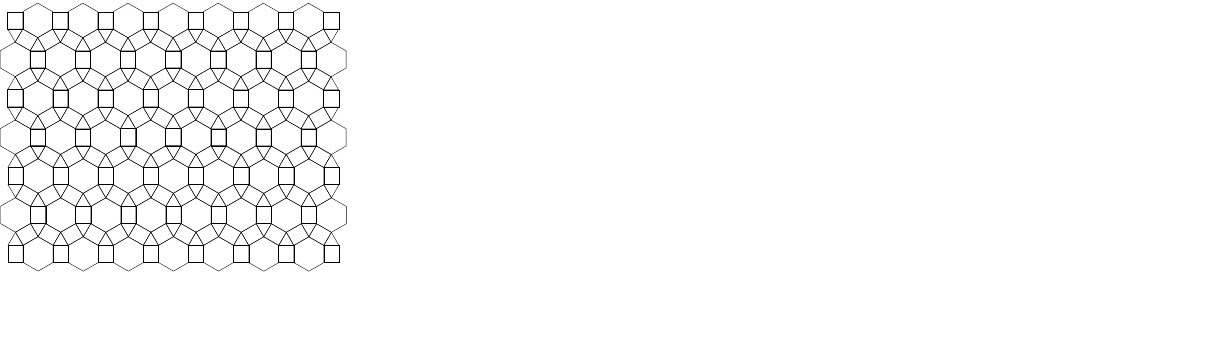}}%
    \put(0.14739223,0.00415473){\color[rgb]{0,0,0}\makebox(0,0)[lt]{\lineheight{0}\smash{\begin{tabular}[t]{l}(a)\end{tabular}}}}%
    \put(0,0){\includegraphics[width=\unitlength,page=2]{triVertex.pdf}}%
    \put(0.50205022,0.00652554){\color[rgb]{0,0,0}\makebox(0,0)[lt]{\lineheight{0}\smash{\begin{tabular}[t]{l}(b)\end{tabular}}}}%
    \put(0,0){\includegraphics[width=\unitlength,page=3]{triVertex.pdf}}%
    \put(0.84454993,0.00870859){\color[rgb]{0,0,0}\makebox(0,0)[lt]{\lineheight{0}\smash{\begin{tabular}[t]{l}(c)\end{tabular}}}}%
    \put(0,0){\includegraphics[width=\unitlength,page=4]{triVertex.pdf}}%
  \end{picture}%
\endgroup%

\caption{Large multi-vertex component.
(a)  Some seed (gray), meet-points (black). Strict Vorono\"{i} cells are limited to the seed themselves (b) closure$^V($merge$(x^{t_c}))$ (black) separates  vertice in strict Vorono\"{i} cells, plus (c) a zone including two multi-vertice (black) equidistant to six seeds.}
\label{fig:triVertex} \end{figure}

\paragraph{Proof.} The circuit's radius is 4 because meet$^E$'s radius is 3. The circuit is shown in fig~\ref{fig:circuitVoronoi} in a folded representation. It allows also to see the radius of computed field: it is the hop count from the gate computing it, to the flip-flop noted MEM storing the boolV layer $x$.
  Function $x\mapsto$merge$(x)$ costs 22+19=41 gates. The total gate count  is 8+1+1+4+41 =55.   Notice first that the vertex-VD is preserved from one iteration to the next: vertex-VD$(x^{t+1})=$ vertex-VD$(x^t)$: %We consider separately two parts of it:
\begin{enumerate}
  \item The part not adjacent to $(x^t)$ is preserved because growth is uniform.
  \item The part adjacent to  $(x^t)$  is precisely  closure$^V($merge$(x^t))$, which means it is directly detected as part of the vertex-VD at time $t$, and will remain empty.
\end{enumerate}
 The configuration $(x^t)_{t\in N}$ is increasing, and will therefore converge at a  time $t_c$.  When this happen, $(x^{t_c})$ is surrounded by closure$^V($merge$(x^{t_c}))$, otherwise $(x^{t_c+1})$ would keep growing. 
 The connected components of vertice within  the complement of closure$^V($merge$(x^{t_c}))$ are of two types: either totally marked or  totally unmarked. The empty components contain some vertices equidistant to three seeds or more. 
 We call those ``{\it multi-vertice}''.  The examples in fig.~\ref{fig:VD} were chosen to illustrate a multi-vertex component consisting of a ball of radius 1, centered on a vertex equidistant to three seeds. As fig.~\ref{fig:triVertex} shows, a  multi-vertex component  can become arbitrary big if we choose a set of seeds regularly spaced on a big discrete circle, so that the circle's center is equidistant to arbitrary many seeds. For the discrete VD, such situations must be considered because they can occur with a non-zero probability. 
  In order to prove that $x^{t_c}$ identifies exactly all the strict Vorono\"{i} cells, we must still prove that the multi-vertex components do not intersect strict Vorono\"{i}-Cell. Let $M$ be one such multi-vertex component. The vertice adjacent to $M$, outside $M$,  are connected because   the graph is planar and maximal. They form a closed curve ${\cal C}(M)$ around $M$ included in closure$^V($merge$(x^{t_c}))$, which is itself included in the vertex-VD.
  A strict Vorono\"{i} cell is connected, and contains a seed, while $M$ does not contain seeds. If one strict Vorono\"{i} cell $V$ was intersecting $M$, let $v$ be a vertex in the intersection, we can apply the Jordan theorem: a path between  the seed of $V$ and $v$, within $V$, would have to intersect the closed curve ${\cal C}(M)$ which is absurd; because ${\cal C}(M)$  is in the vertex-VD which is the complement of strict Vorono\"{i} Cell.

\section{Conclusion} 
\paragraph{Contrasting Cellular Circuits with CA.}
The usual scheme for specifying  cellular computation, Cellular Automata (CA), uses a lattice of Processing Elements (PE) exchanging their finite state between direct neighbors, and applying a Look-up Table (LUT) to find out the next state. 
In this paper, we   present  a new scheme using spatial types and producing cellular circuits. The goal is to explore the world of cellular computation beyond lattice networks to reach amorphous computing, and along the complexity axis of elaborate rules with high radius.

It is not automata that are mapped on  a network's vertices, but bits and gates. The network can be any maximal planar graph.
% although we illustrate examples with the hexagonal lattice, for convenience.
  Bits and gates are distributed not only on vertices, but also on edges and faces of this graph, and also on secondary locus in-between those. No LUT are used.
Instead, fields of those bits are computed from other fields using 
  functions programmed with spatial operations. Those operations
    let data travel between adjacent vertices,  edges and faces, and interact through reductions.
Operation-expressions can be  directly translated into    circuits of logical gates. The new scheme improves efficiency and programmability.%, and this enables a complex application: implementing an SDN-medium.

\subsection{Improving efficiency}

\paragraph{Exploiting symmetries to factorize computation.} Cellular circuits  exploit the  spatial symmetries always occurring when doing artificial physics. For example, let  $x$ be a boolV (a boolean vertex field). Consider the computation of  $x\mapsto$ frontier$^E(x)$ which returns true for edges on the frontier of an $x$-blob. It is  symmetric  with respect to the edge's  adjacent vertices $v_1,v_2$: one must be empty, and the other filled. In a PE+LUT scheme, the computation of  $x\mapsto$frontier$^E(x)$ must be done on PEs assigned to vertices. It has to be computed two times for $v_1$ and $v_2$, and stored as a vector of 6 boolV, (bit density of 6) with no obvious visualization. 
With spatial types, it is computed a single time, using a xor reduction. It is stored as a unique boolean edge field (a boolE with bit density of 3). There is an automatic  nice visualization looking like a set of closed curves around each blob (fig.~\ref{fig:feature}), which   precisely remind frontiers. 
The speed up brough by exploiting symmetries with reduction really makes a difference when several reduction are applied in sequence, generating high radius field. It results in a change of time complexity  with respect to the radius, as we now detail:

  \paragraph{Translating cellular circuits into CAs.} If the planar-graph is a lattice, the cellular circuit can be tiled with the exact same building block circuit (as is done in fig.~\ref{fig:growCircuit} for the degenerated 1D case).
The cellular circuit can be translated into a formal CA, by  doing the computation of each tile on the  PE of a CA.
However, as fig.~\ref{fig:growCircuit} shows, the tiles exchange not only the stored state, but also intermediate values that have already used the neighbor state in their computation. 
 In contrast, in  CAs, each PE   receives only the stored state from neighbor.
% On top of the computation done by its associated tile, 
In the CA translation, each PE must therefore  redo part of the computation done by neighboring tiles. The analysis done in subsection~\ref{subsection:complexity} shows that for a computed field $f$ of radius $r$, this extra work causes   the time complexity to jump from   $O(r)$ to   $O(r^3)$ making it unfeasible for large $r$ (for our current  SDN medium $r=25$).

%When doing articifial physics,  symmetry  is the rule, not the exception. 

    %: we used a radius of 25 in~\cite{utubeDevelopement2018}.
 
% 
%  The GPCA needs to
%    This means exploiting all the repertoire of the possibilities offered by the CA, and in particular, programming it as a dynamical system made of many interacting sub-layers.
%    
%     
% Problems to solve with a CA, where a highly  complex transition rule is needed, simply do not occur in the CA litterature.
% The famous self-reproducing CA of Von Neumann, with 29 states~\cite{Neumann:1966} is to our knwoledge, the sole outstanding exception to this statement. 
% But event in the theme of selfreproduction, solution  using  simple CA rules  were presented later on, such as the Langton's loop~\cite{langton1984self}. 
%  
% One can ask this simple question: why has a complex transition never been really needed?
% Our answer is: because 
%  The PE + LUT approach is fit for immediate or at least, next-to-immediate, radius-2 neighborhood.
%   On the contrary, cellular-circuits handle effortlessly high radius,  because computation and communication are interleaved at the finest possible granularity: the bit level.
%   As soon as a new intermediate boolean field is computed throughout the cellular space, its values can be sent again to neighbors, thereby incrementing the radius. 
%    

\paragraph{SIMD pipelined execution.}   We used exclusively the hexagonal lattice with 64 columns for simulating the SDN-medium. Our simulator process the lattice row by row, in a pipeline way. It has two advantages: 1- it exploit the SIMD capability of standard PC: 64 logic gates can be evaluated in a single logic operation on long integer, 64 bits can be communicated with a single bit rotation\footnote{We measured more than 64 gate evaluation per clock-cycle, due to the super-scalar capabilities  present in standard laptops}.2- Rows of generated intermediate fields are consumed at the same rate as they are created. As a result, only $r$ rows  have to be stored, where $r$ is the radius.

% We choose this problem for several reasons: 
% 1-  It can be done with traditionnal CA using few states, but then then the look up table has still to be programmed and that is not trivial.
% 2-  In contrast, we solve it using  only a few spatial operations.
% Furthemore we  compute intermediate fields which have a meaning, and can also be visualized.
% 3- The presentation of our model fills the paper, and there is no room for reporting a more complex example, the execution of more complex example can be viewed in three utubes~\cite{utubeDevelopement2018}.  ;
 %Allowing to use both edge and vertices facilitates the formulation:

 \subsection{Improving programmability  through procedural programming.}
  This is probably the most significant advantage of spatial types. When programming a cellular circuit, we do not directly focus on achieving a specific update function. Instead, we  program and debug separately  a library of functions, that we reuse later, just as we do with standard procedural programming languages. 
  For example in this paper, we introduce first low-level functions to compute blob features such as the inside, the outside, the frontier, the neighborhood. We also implement a one-to-one communication-operation between apex neigbors as a function. Those functions have radius 1  or 2, and need less than 10 gates.   Let $x$ be a boolV representing the support of agents.  We then program the more complex function of radius 3: $x\mapsto$merge$(x)$ 41 gates (resp. $x\mapsto$divide$(x)$) used for moving agents without merging (resp. without dividing) their support.
The disjunction $x\mapsto$meet$(x)=$merge$(x)\vee$divide$(x)$ is  is a key building block reused dozens of time in the SDN-media in order  to preserve agent's support. 
  In this paper we reuse the $x\mapsto$merge$(x)$ component   adding 14  gates, in order to obtain a radius-4, 55-gates cellular-circuit  computing the VD which is not  related to the SDN-media\footnote{An SDN-medium does compute a VD, but a dynamic one, using a much more complex circuit: it computes distances modulo 8 (~\cite{integerGradient}), and constantly updates the VD as the seeds are moving simultaneously with their VD computation.}: the discrete Vorono\"{i} Diagram.

  \paragraph{Using  auxiliary fields.}
CA transition rules tends to be simple. One factor limiting their complexity is that during an update cycle, there is only a limited set of data availabe as inputs: the local state plus the states of the other  PEs in the immediate neighborhood.  
With spatial types, one update cycle includes  an arbitrary number of $\mu$steps of exchange-compute. Each $\mu$step is a reduction which produces  a new auxiliary field, that can be exchanged again, to become the input of another $\mu$steps.
 The generated auxiliary fields increase the volume of data on which it is possible to compute. They are like  auxiliary variables used in procedural programing: they store intermediate results   to be re-used several times for different purposes. 
 For example the auxiliary field frontier$^E(x)$ appears in formula~\ref{eq:meetE} and is  reused   for computing nbcc$(x)$  and frontier$^V(x)$. So it is reused three times in total for the computation of $x\mapsto$ meet$(x)$. 
Designing increasingly complex transitions naturally results from  adding new $\mu$steps, and is most often done without having to introduce more bits of state.
An illustration of this fact is that in this paper, the discrete Vorono\"{i} diagram  is computed with a single bit of state $x$. 
%COMPARAISON 

%Each time an auxiliary field is exchanged, its   radius is incremented.  In practice, the radius tends to grow  linearly with the operation-expression depth, because computation and communication are interleaved at the finest possible granularity: the bit level.
% Cellular-circuits handle effortlessly high radius, because their complexity augments only proportionaly to the radius. 
 
\subsection{Application: from the Voronoi Diagram to the SDN medium} 

\paragraph{The discrete Vorno\"{i} Diagram (VD).} We choose this example because 1- it was simple enough to fit in the paper, and 2- it reuses a key function needed for SDN media, allowing us to start presenting it 3- It has also an intrinsic interest. 
   Computing the VD on a CA is not new, we use a technique inspired from~\cite{Adamatzky1996,VDwave}: waves   propagate synchronously and define the VD when they collide. Spatial types can  do it for the more  general context of maximal planar graph. The program can capture the simple algorithmic essence of the wave technique, which is  to grow the seeds uniformly as much as possible and stop just before they meet.   
  The circuit uses  55 gates on each tile, for the hexagonal case. We conjecture that it is the minimum.   The number 55 measures the complexity of the computation in a  more precise way than just the  number of states  which is traditionally used  in the CA community.
  There is only one bit of state, compared to two bits for~\cite{Adamatzky1996} (four states) and $0(ln(n))$ bits for~\cite{VDwave}, where $n$ is the number of PEs. 
    As in amorphous computing, the hypothesis of synchronism is not mandatory for the   circuit's operation. If the unique bit of state  is not updated with a small probability, uniformly on each PE, then  the  circuit  will still compute  an approximation of the VD, with fluctuation due to variation in the propagation speed.

\paragraph{The SDN medium} Spatial types and cellular circuits  were developed as a necessary tool needed to construct piece by piece a quite complex computing medium which can simulate Self Developing Network (SDN). The goal of SND-media is to broaden the scope of what can be computed on a computing medium, and reach general purpose computing. 
   The complexity of SND-media is due to the simulation of many artificial physical laws, needed to achieve division of homogenized membranes.  
   %Spatial types are specially  fit for working with geometric 2D features, which  is precisely what is needed for the GPCA.  
   Our current version, uses 77 bits of state 14,291 gates,  300 trans-wires between cells and  a radius of 25.
   A real-time execution, interpreting a flow of host-instructions dictating  the  development of a virtual 2D-grid  can be viewed on the   three videos~\cite{utubeDevelopement2018}, it gives  an idea of how complex computational behavior can be obtained thanks to the new scheme, while still doing bit-level local communication, the essence of CA.

\section*{Acknowledgment} We thank Luidnel Maignan for his fruitful comments.
% 
% \section*{Compliance with Ethical Standards}
%  Author Frederic Gruau declares that he has no conflict of interest.
%   This article does not contain any studies with human participants or animals performed by any of the authors.

 \bibliographystyle{splncs04} 
 % Hack natbib so it matches the LNCS style: reference list in a
% section with small font and no square brackets.
\renewcommand\bibsection
  {\section*{\refname}\small\renewcommand\bibnumfmt[1]{##1.}}
  
%\bibliography{../common/HDR}

\section*{Appendix}
 We here report more technical issues.

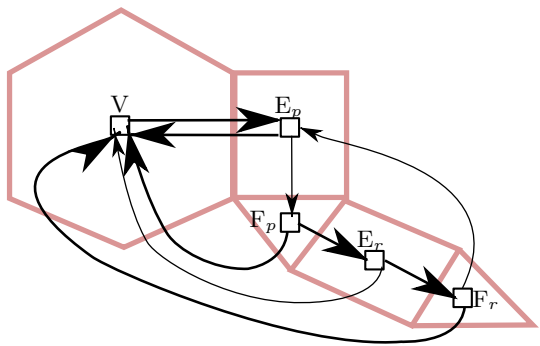
\begin{figure} \centering   \def\svgwidth{7cm }   
%% Accompanies image file 'radius.pdf' (pdf, eps, ps)
\begingroup%
  \makeatletter%
  \providecommand\color[2][]{%
    \errmessage{(Inkscape) Color is used for the text in Inkscape, but the package 'color.sty' is not loaded}%
    \renewcommand\color[2][]{}%
  }%
  \providecommand\transparent[1]{%
    \errmessage{(Inkscape) Transparency is used (non-zero) for the text in Inkscape, but the package 'transparent.sty' is not loaded}%
    \renewcommand\transparent[1]{}%
  }%
  \providecommand\rotatebox[2]{#2}%
  \newcommand*\fsize{\dimexpr\f@size pt\relax}%
  \newcommand*\lineheight[1]{\fontsize{\fsize}{#1\fsize}\selectfont}%
  \ifx\svgwidth\undefined%
    \setlength{\unitlength}{286.43125665bp}%
    \ifx\svgscale\undefined%
      \relax%
    \else%
      \setlength{\unitlength}{\unitlength * \real{\svgscale}}%
    \fi%
  \else%
    \setlength{\unitlength}{\svgwidth}%
  \fi%
  \global\let\svgwidth\undefined%
  \global\let\svgscale\undefined%
  \makeatother%
  \begin{picture}(1,0.63742167)%
    \lineheight{1}%
    \setlength\tabcolsep{0pt}%
    \put(0,0){\includegraphics[width=\unitlength,page=1]{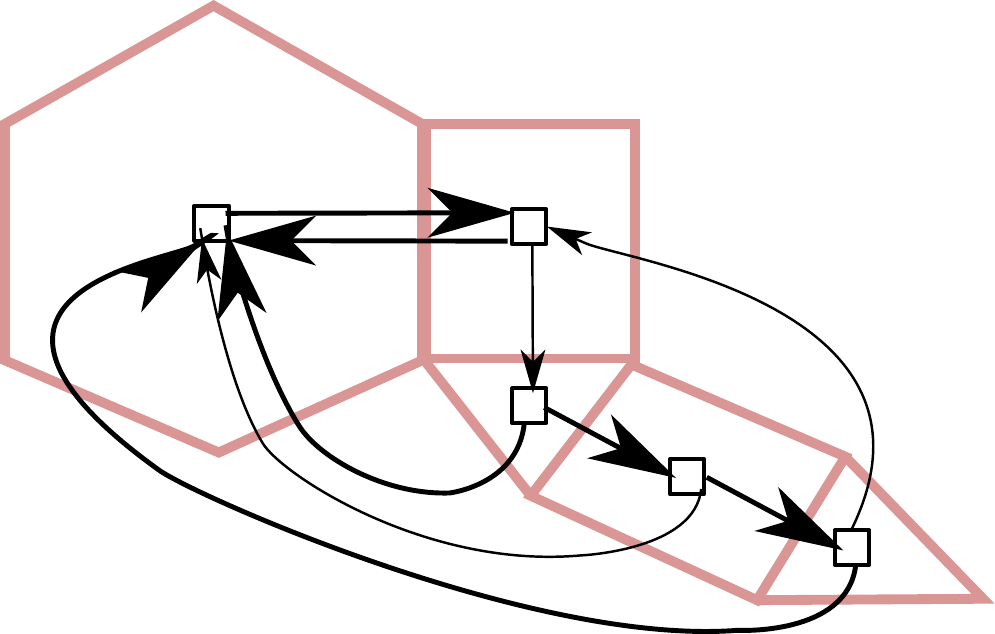}}%
    \put(0.19472348,0.43799578){\color[rgb]{0,0,0}\makebox(0,0)[lt]{\lineheight{1.25}\smash{\begin{tabular}[t]{l}V\end{tabular}}}}%
    \put(0.502838,0.43612877){\color[rgb]{0,0,0}\makebox(0,0)[lt]{\lineheight{1.25}\smash{\begin{tabular}[t]{l}E$_p$\end{tabular}}}}%
    \put(0.65795321,0.18277541){\color[rgb]{0,0,0}\makebox(0,0)[lt]{\lineheight{1.25}\smash{\begin{tabular}[t]{l}E$_r$\end{tabular}}}}%
    \put(0.45621018,0.22083835){\color[rgb]{0,0,0}\makebox(0,0)[lt]{\lineheight{1.25}\smash{\begin{tabular}[t]{l}F$_p$\end{tabular}}}}%
    \put(0.874932,0.07024683){\color[rgb]{0,0,0}\makebox(0,0)[lt]{\lineheight{1.25}\smash{\begin{tabular}[t]{l}F$_r$\end{tabular}}}}%
  \end{picture}%
\endgroup%

\caption{Finite State automaton for computing the radius: state distinguishes perimeter-edges and perimeter-faces noted $E_p,F_p$, from radius-edges or radius-faces noted $E_r,F_r$. Transitions correspond to transfer communications.  Thick transitions indicates when the radius is incremented.}
\label{fig:radius} \end{figure}

 \paragraph{Computing the radius of a circuit.}
 The radius of layers is 0, since they are directly read from memory. 
 Usually the radius is incremented, each time a transfer is done. However, fig~\ref{fig:circuitVoronoi} shows a conter example: the field $x\mapsto$nbcc$(x)$ undergoes three transfers: from vertex to edge to face to vertex again, but its radius is only 2, (it is clearly computed   on vertices around the starting vertice of reference). 
 The second transfer, from edges to faces did not increase the radius, because the faces adjacent to the edges at distance 1 from the starting vertice are also at distance $\leq 1$. In the same way  edges (resp vertices) which are adjacent to faces (resp. edges) at distance $2$, are also at distance $\leq 2$. We call  edges (or faces) at distance 1 (resp. 2) ``{\bf perimeter}''   (resp. ``{\bf radial}'') edges (or faces).
 When computing the radius of an auxiliary field, by induction on the operator expression, we must remember for edges or faces, whether it is a perimeter, or a radial one. We take this ``sub-type''   into account when a transfer occurs, to know wether the radius is incremented or not. Fig~\ref{fig:radius} shows the finite state automaton that does this job.  
 What matters for the computation is the parity of transfers done since the last vertex. If communications is done only between edges and faces, the radius augments on average only one transfer out of two. 
 If a binary non-spatial operation is applied to two fields, then the resulting radius is of course, the max of the radius of the input fields, and if the locus is edges or faces, we have radial$>$perimeter when determining the resulting sub-type. 
 
\begin{figure} \centering  
\def\svgwidth{\columnwidth} 
  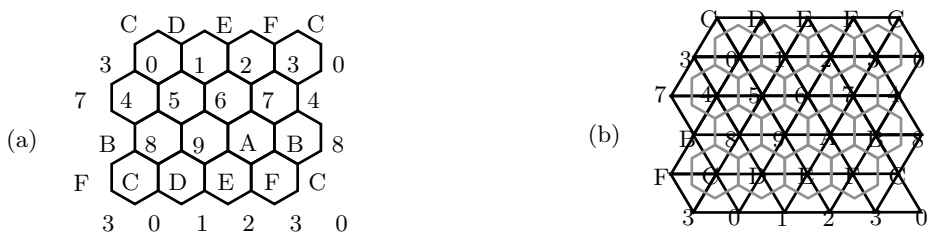
\caption{The 4 $\times 4$ hexagonal lattice: (a) The toroidal wrapping. (b) The underlying planar graph  (c)Voronoi polygons associated to V, E, F data points.  } \label{fig:hexLat} \end{figure}

\paragraph{Processing of the border.}
A circuit is a finite object. 
On a 2D plane,  a difficulty occurs on the border of the planar graph: the unbounded face is not triangular: it is adjacent to all the vertices on the perimeter which number is $O\sqrt(n)$ where $n$ is the total number of vertices. There are two solutions:
1-The simplest  is to come back to a triangulated form by considering a 2D-torus instead of a 2D-plane. 
Simplicial proximity can also be defined on 
  toroidal-graph , whose vertices are mapped on a 2D torus, and edges do not cross.  
In our experiments, we use the perfect  hexagonal toroidal graph shown in fig~\ref{fig:hexLat}~(a) (with 64 columns). The same construction can be done in the isotrope case. 
2- Ultimately, the border will have to be instanciated, so as to model input and output to the cellular circuit. 
Notice that the vertices of the border form  a 1D-ring. This ring is not used
for computation. When a reduction of a vertex field is done, the  field is prolongated on the vertice of the ring, by setting the value to the neutral element of the reduction.
In the example of the Voronoi Diagram, frontier$^E$ is computed using a xor reduction, and the neutral element for xor is zero. It means that the seeds on the border behave as if they were surounded by empty vertices. 

\end{document}